# Emergence of large spin-charge interconversion at an oxidized Cu/W interface

Inge Groen[1], Van Tuong Pham[2], Stefan Ilić[3], Won Young Choi[1], Andrey Chuvilin[1,4], Edurne Sagasta[1], Diogo C. Vaz[1], Isabel C. Arango[1], Nerea Ontoso[1], F. Sebastian Bergeret[3,5], Luis E. Hueso[1,4], Ilya V. Tokatly[4,5,6], & Fèlix Casanova[1,4].

[1] CIC nanoGUNE BRTA, 20018 Donostia-San Sebastian, Basque Country, Spain.
[2] Université Grenoble Alpes, CNRS, Institut Néel, 38042 Grenoble, France.
[3] Centro de Física de Materiales (CFM-MPC), Centro Mixto CSIC-UPV/EHU, 20018 Donostia-San Sebastian, Basque Country, Spain.
[4] IKERBASQUE, Basque Foundation for Science, 48009 Bilbao, Basque Country, Spain.
[5] Donostia International Physics Center (DIPC), 20018 Donostia-San Sebastian, Basque Country, Spain.
[6] Nano-Bio Spectroscopy Group and European Theoretical Spectroscopy Facility (ETSF), Departamento de Polímeros y Materiales Avanzados: Física, Química y Tecnología, Universidad del País Vasco, 20018 Donostia-San Sebastián, Basque Country, Spain.

## Abstract

Spin-orbitronic devices can integrate memory and logic by exploiting spin-charge interconversion (SCI), which is optimized by design and materials selection. In these devices, such as the magnetoelectric spin-orbit (MESO) logic, interfaces are crucial elements as they can prohibit or promote spin flow in a device as well as possess spin-orbit coupling resulting in interfacial SCI. Here, we study the origin of SCI in a Py/Cu/W lateral spin valve and quantify its efficiency. An exhaustive characterization of the interface between Cu and W electrodes uncovers the presence of an oxidized layer ($WO_x$). We determine that the SCI occurs at the Cu/$WO_x$ interface with a temperature-independent interfacial spin-loss conductance of $G_\parallel \approx 20 \times 10^{13}\ \Omega^{-1} m^{-2}$ and an interfacial spin-charge conductivity $\sigma_{SC} = -1610\ \Omega^{-1} cm^{-1}$ at 10 K ($-830\ \Omega^{-1} cm^{-1}$ at 300 K). This corresponds to an efficiency given by the inverse Edelstein length $\lambda_{IEE} = -0.76$ nm at 10 K ($-0.4$ nm at 300 K), which is remarkably larger than in metal/metal and metal/oxide interfaces and bulk heavy metals. The large SCI efficiency at such an oxidized interface is a promising candidate for the magnetic readout in MESO logic devices.

## I. INTRODUCTION

In the field of spin-orbitronics great efforts have been made to develop theories and device architectures that allow technological advancements in digital information storage and computation. These developments include writing and reading of information in magnetic states to achieve scalable and energy-efficient spin-orbit memory such as spin-orbit torque magnetoresistive random access memory (MRAM) [1,2]. Now, the field is pushing even further with proposals for spin-orbit logic [3]. These devices rely on spin-orbit coupling (SOC) which permits spin-charge interconversion (SCI), that is, the conversion of a spin current density into a charge current density or vice versa. Such conversion can take place in bulk materials (3D) via the spin Hall effect [4] and at interfaces (2D) via the Rasbha-Edelstein effect [5–12], and in specific systems, both bulk and interfacial SCI are observed [13,14]. Spin-orbit devices benefit from spin-orbit materials (SOMs) providing high SCI efficiencies. For optimization of these efficiencies, it is important to distinguish between bulk and interfacial SCI which is not straightforward since interfaces are present in most devices.

Lateral spin valve (LSV) devices can be used for SCI measurements in both 2D and 3D systems [10,12,15,16]. The non-local geometry of this device allows separating pure spin current from charge current such that spurious effects arising from the latter are eliminated from the detected signal. Two independent configurations help to determine the spin diffusion length ($\lambda$) and the SCI efficiency of a SOM. However, the spin transport at the interface of the channel and SOM electrode remains an issue. Lately, the importance of interfaces has been pointed out by several works [17–20] where the interface can reduce or even prevent spin sink into SOM, or possess 2D SCI due to interfacial SOC. Hence, the quantification of the 3D SCI in the SOM might be affected by the presence of an interface.

Here, we study SCI in an all-metallic Py/Cu/W LSV. Electrical measurements of the Cu/W interface supported by scanning transmission electron microscopy (STEM) images reveal that the interface resistance originates from an oxide layer created during the fabrication of the device. After analyses of different spin transport scenarios in the system, we identified that the SCI occurs at the Cu/oxide interface. We find a temperature-independent interfacial spin-loss conductance ($G_\parallel \approx 20 \times 10^{13}\ \Omega^{-1}\ m^{-2}$) and a spin-charge conductivity of $\sigma_{SC} = -830\ \Omega^{-1}\ cm^{-1}$ at 300 K, corresponding to an inverse Edelstein length $\lambda_{IEE} = -0.4$ nm at 300 K for the Cu/oxide interface. The high resistivity and large SCI in this system is promising for developing the magnetic-state readout in the magnetoelectric spin-orbit (MESO) logic device [21,22]. This work highlights the importance of studying every aspect of all-metallic devices carefully, in particular the interface, to extract meaningful spin relaxation and SCI efficiency parameters.

## II. EXPERIMENTAL DETAILS

The Py/Cu/W LSV consists of two parallelly aligned Py electrodes and a W electrode placed in between the Py electrodes. These three electrodes are connected by a transverse Cu channel. Figure 1(a) displays a top-view scanning electron microscopy (SEM) image of the device. The devices are fabricated on SiO$_2$(150 nm)/Si substrates in three steps, each step involving electron-beam lithography (EBL), metal deposition, and a lift-off process. The first step includes EBL of the Py nanostructure, followed by Py deposition via e-beam evaporation (rate $\sim 0.56\ \text{Ås}^{-1}$ and $p_{dep} \sim 2.3 \times 10^{-8}$ Torr). In a second EBL step, the nanostructure for the W is defined with subsequent deposition of W by magnetron sputtering (rate $\sim 0.11\ \text{Ås}^{-1}$, $p_{Ar} = 3$ mTorr, $P = 10$ W, $p_{base} \sim 2 \times 10^{-8}$ Torr at room temperature). After lift-off, Ar-ion-beam milling is performed at grazing incidence to remove sharp vertical edges from the Py and W electrodes formed by materials deposited on the walls of the resists which did not detach from the electrodes during the lift-off process. Lastly, the Cu nanostructures are defined by EBL. Before the deposition of Cu, Ar-ion-beam milling is performed at normal incidence to clean the Py and W surfaces and 3 nm of Cu is deposited in-situ by magnetron sputtering (rate $\sim 1.88\ \text{Ås}^{-1}$, $p_{Ar} = 3$ mTorr, $P = 250$ W, $p_{base} \sim 3 \times 10^{-6}$ Torr at room temperature) in an attempt to create a highly transparent interface between the transverse Cu channel and the Py and W electrodes. Then, the rest of the Cu is deposited by thermal evaporation in a different chamber (rate $\sim 1.5\ \text{Ås}^{-1}$ and $p_{dep} \sim 1.3 \times 10^{-8}$ Torr) and subsequently lifted off in acetone. Finally, the sample is capped with a magnetron sputtered 5-nm-thick film of SiO$_2$ (rate $\sim 0.5\ \text{Ås}^{-1}$, $p_{Ar} = 3$ mTorr, $P = 200$ W, $p_{base} \sim 2 \times 10^{-8}$ Torr at room temperature).

The width ($w$) and thickness ($t$) of the electrodes for the device used in this study are $w_{Py} = 124$ nm, $t_{Py} = 30$ nm, $w_W = 195$ nm, $t_W = 4.5$ nm, $w_{Cu} = 123$ nm and $t_{Cu} = 90$ nm. The measured resistivities of the different electrodes are presented in Supplemental Material S1 [23]. The behavior of the W resistivity indicates that the $\beta$-phase, which possesses large SCI, is dominant [24]. The transport measurements are carried out in a physical property measurement system (PPMS) from Quantum Design. We can apply an in-plane magnetic field and the electrical measurements are performed using the "dc reversal technique" with a Keithley 2182 nanovoltmeter and 6221 current source.

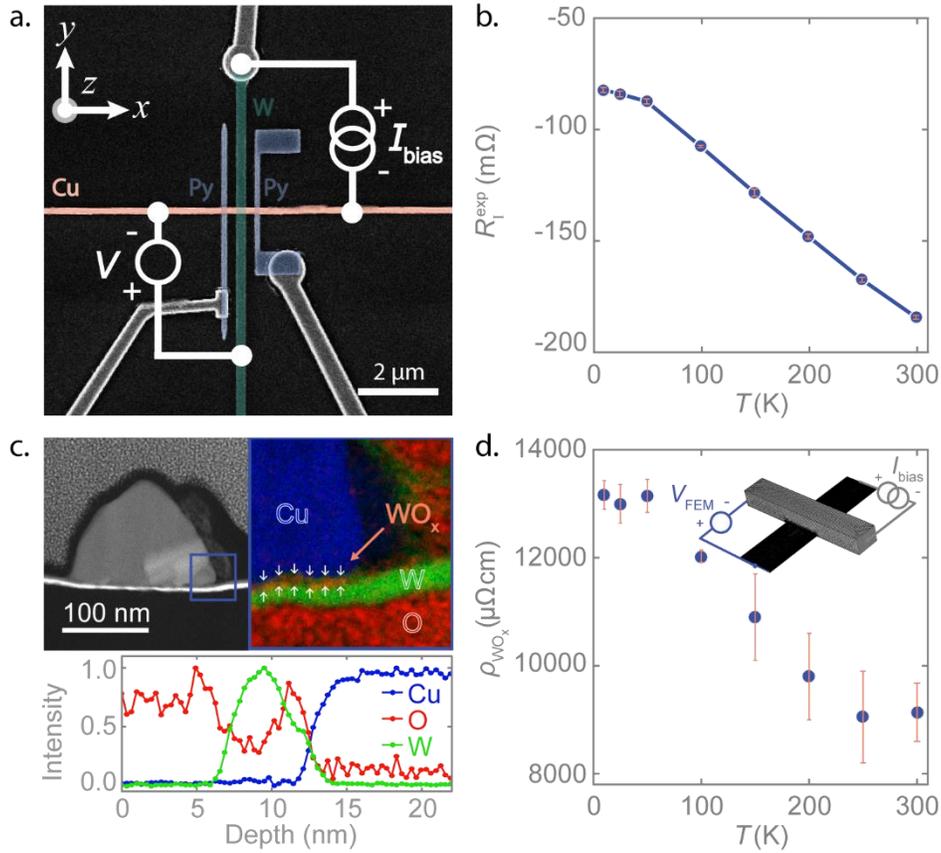

FIG. 1: Characterization of the Cu/W interface. (a) A false-color SEM image of a Py/Cu/W LSV with the Py, Cu and W electrodes indicated by blue, orange, and green, respectively. The electrical configuration presents the 4-probe interface resistance measurement of the Cu/W interface. (b) Temperature dependence of the experimental interface resistance $R_I^{exp}$ of the Cu/W interface. (c) A STEM image and EDX analysis of the Cu/W interface. Top left panel: HAADF STEM image of a cross-section of the Cu/W interface area. Top right panel: False-colored EXD map of the blue square on the STEM image. The false colors correspond to O (red), Cu (blue), and W (green). An interfacial oxide layer is indicated by white arrows. Bottom panel: Compositional analysis of the EDX map showing the Cu, O and W as a function of the depth, indicating the presence of a $WO_x$ layer in between the Cu and W electrodes. (d) The resistivity of the $WO_x$ layer as a function of the temperature, extracted from the 3D FEM simulation in combination with $R_I^{exp}$. Inset: the geometrical model and measurement configuration in the 3D FEM simulation.

### III. RESULTS and DISCUSSION

#### A. Interface resistance

The 4-point measurement configuration in the LSV device shown in Fig. 1(a) can be used for probing the experimental interface resistance $R_I^{exp}$. Figure 1(b) shows the resulting interface resistance at various temperatures. $R_I^{exp}$ is observed to be negative indicating a low-impedance interface. The negative value for interface resistance is an artifact that comes from an inhomogeneous current density flow and voltage drop facilitated by the low-impedance interface [25,26]. 3D finite element method (FEM) simulations are utilized to extract the actual interface resistance.

STEM has been performed to investigate the W and Cu channels at the interface region and to accordingly define the geometrical model within the 3D FEM resembling the Cu and W electrodes. Figure 1(c) shows a cross-sectional STEM image of the Cu/W interface. The element composition is inspected by energy dispersive x-ray (EDX) analysis in the area indicated by the blue square. The observed elements in this region are oxygen (O), W and Cu. Surprisingly, O appears in between the Cu and W electrodes. Supplemental Material S2 contains a deeper discussion on the elemental composition [23]. We find the W electrode thickness below the Cu channel to be 2.8 nm thick. The oxide layer, clearly visible in the zoom of the blue square indicated by white arrows, has an average thickness of ∼1.5 nm and consists of W and O. In the remainder of this paper, we will refer to this layer as oxide layer and use the tag "$WO_x$".

The finite thickness of the oxide layer means that the interface in the 3D FEM simulation is not to be considered as a boundary condition, but the geometrical model must contain an additional layer with its own materials properties. The geometry has been constructed considering the structural details found in the TEM images (Supplemental Material S2) [23]. Although the thickness of the W electrode underneath the Cu channel and away from the Cu channel is different, we consider the resistivity of the W electrode to be the same everywhere. The electrical configuration in the simulation is shown in the inset of Fig. 1(d) [and is the same as presented in Fig. 1(a)]. The voltage is probed while varying the resistivity of the oxide layer $\rho_{WO_x}$. The resistance for which the FEM resistance $R_{FEM}$ is equal to $R_I^{exp}$ gives the correct $\rho_{WO_x}$ (see Supplemental Material 0 for more details [23]). Note that the described method can also be used when there is no interfacial oxide layer by replacing the finite layer by a contact impedance and the resistivity by an interface resistance. Figure 1(d) graphs the resulting temperature dependence of $\rho_{WO_x}$ which is relatively high compared to resistivities in the metal electrodes (Supplemental Material S1 [23]) and is considered in the following studies on the spin absorption and SCI.

### B. Spin absorption

The spin absorption technique is used for defining $\lambda$ of materials such as ferromagnets and SOM with short $\lambda$. Figure 2(a) displays the SEM image of our LSV with the measurement configuration for the spin absorption technique. A bias current $I_{bias}$ is applied from the Cu channel into the Py electrode creating a spin accumulation at the Py/Cu interface. Spin diffusion in the Cu channel away from the interface region results in a pure spin current (on the side where $I_{bias}$ does not flow). A second Py electrode is used to detect this spin current. The W electrode positioned in between the two Py electrodes can absorb part of the spin current flowing in the Cu channel depending on spin properties of the W, the $WO_x$ layer and the W/$WO_x$ and Cu/$WO_x$ interfaces.

Figure 2(b) plots the magnetic field dependence of the non-local resistance ($R_{NL} = V_{NL}/I_{bias}$) in a reference LSV without the W electrode ($R_{NL}^{ref}$) and the LSV with the W electrode ($R_{NL}^{abs}$) at 10 K. The two Py electrodes are designed in such a way that the magnetic switching fields are different. Therefore, the parallel and antiparallel resistance states can be identified by parallelly aligning the ferromagnetic metal (FM) with an external magnetic field and then sweeping this field [15,16]. The difference between the resistance of the parallel and antiparallel configuration are the spin signals $\Delta R_{NL}^{ref}$ and $\Delta R_{NL}^{abs}$. Figure 2(b) shows that $\Delta R_{NL}^{abs}$ is smaller than $\Delta R_{NL}^{ref}$, meaning that the W electrode with the interfacial $WO_x$ layer allows for spin absorption. Figure 2(c) presents the temperature dependence of the spin signals. The observed $\Delta R_{NL}^{abs}$ is of the same order of magnitude as other all-metallic LSV based on Nb [15], Pt [16], Ta [27], CuIr [28], CuBi [29] or AuW [30] electrodes or even by an electrode constructed out of metallic/oxide heterostructures [10].

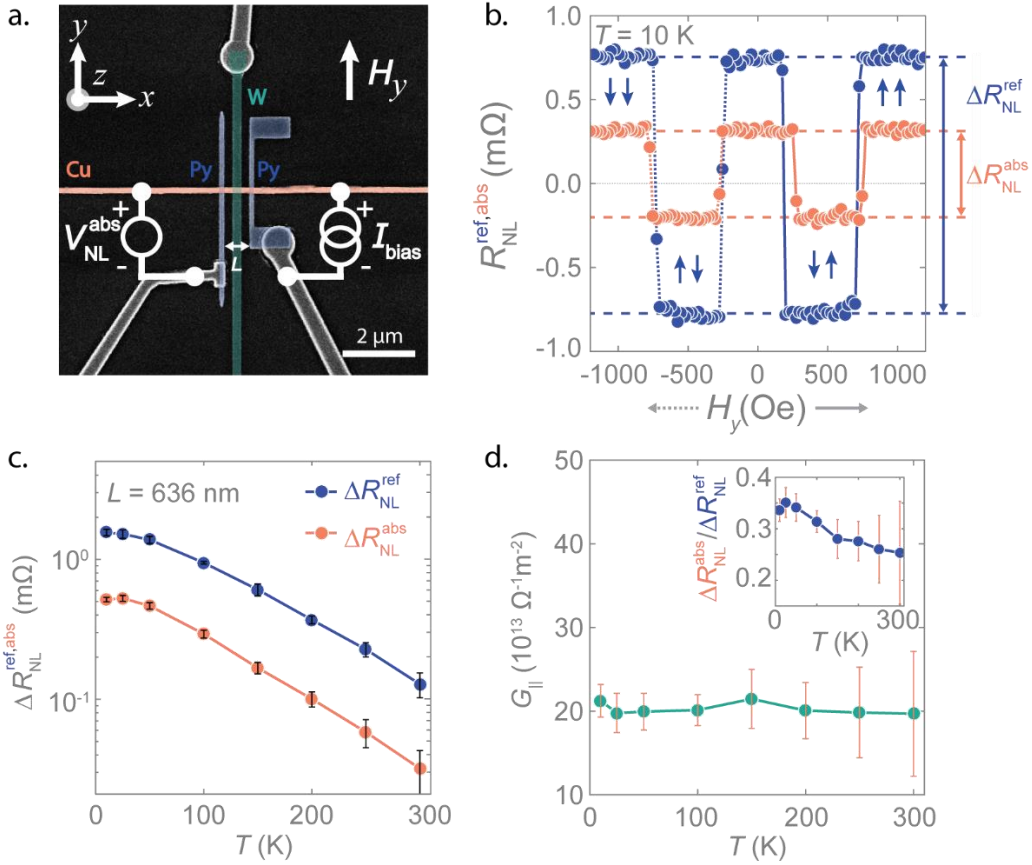

FIG. 2: Spin absorption in the Py/Cu/W LSV. (a) A false-color SEM image presenting a top view of the Py/Cu/W LSV including the spin absorption technique measurement configuration. The magnetic field is applied along the easy axis of the Py electrodes. (b) The non-local resistance for the reference LSV (blue) and the LSV with the middle W electrode (orange) at 10 K. The difference between the low and high resistance state gives the spin signals $\Delta R_{NL}^{ref}$ and $\Delta R_{NL}^{abs}$ for the associated LSVs. (c) The spin signals $\Delta R_{NL}^{ref}$ and $\Delta R_{NL}^{abs}$ at different temperatures. (d) The temperature dependence of the interfacial spin-loss conductance $G_\parallel$ obtained from applying equation (2) to $\Delta R_{NL}^{abs}/\Delta R_{NL}^{ref}$ which is displayed in the inset.

The value of $\lambda$ in a SOM using the spin absorption technique is typically obtained in devices with a transparent interface between the spin transport channel of a non-magnetic metal (NM) and a SOM electrode. However, in LSVs with resistive interfaces, that is, LSVs in which the spin resistance of the interface is dominant over the one of SOM, the spin properties of the SOM cannot be accessed and only the interfacial properties can be explored. Interestingly, spins are being absorbed in our Py/Cu/W LSV even though the Cu/W contains a resistive interfacial oxide layer. The ratio of the spin signals $\eta = \Delta R_{NL}^{abs}/\Delta R_{NL}^{ref}$ is used to pinpoint where the spins are absorbed and elucidate the spin properties that are probed. We assume that the measured interface resistance comes from the oxide resistivity and the W/WO$_x$ and Cu/WO$_x$ interfaces are transparent, as we cannot differentiate the contribution of each candidate to the experimental interface resistance. Then, $\eta$ for bulk and interfacial dominant spin absorption are respectively given by:

$$\eta_{\text{bulk}} = \left[1 + \frac{G_s^{\text{SOM}}}{G_s^{\text{NM}}}\left(\frac{1}{2} - \frac{1}{1 + r_{\text{FM}}e^{L/\lambda_{\text{NM}}}}\right)\right]^{-1}, \quad (1)$$

and

$$\eta_{\text{interfacial}} = \left[1 + \frac{G_\parallel A_\text{I}}{G_\text{s}^{\text{NM}}}\left(\frac{1}{2} - \frac{1}{1 + r_{\text{FM}}e^{L/\lambda_{\text{NM}}}}\right)\right]^{-1}, \quad (2)$$

with $r_{\text{FM}} = 1 + 2G_\text{s}^{\text{NM}}/G_\text{s}^{\text{FM}}$ where the spin conductance for the NM channel $G_\text{s}^{\text{NM}} = A_{\text{NM}}/\rho_{\text{NM}}\lambda_{\text{NM}}$, the FM electrode $G_\text{s}^{\text{FM}} = (1 - P^2)A_{\text{FM}}/\rho_{\text{FM}}\lambda_{\text{FM}}$ and the SOM electrode $G_\text{s}^{\text{SOM}} = A_{\text{SOM}}/\rho_{\text{SOM}}\lambda_{\text{SOM}}$ with $A$ being the cross-sectional area through which the spin current flows in the associate electrode, and $P$ is the polarization of the FM electrode. $G_\parallel$ is the interfacial spin-loss conductance and $A_\text{I}$ is the NM/SOM interface area. The derivation of $\eta$ for the total spin absorption including both bulk and interfacial absorption is shown in Supplemental Material S4 [23]. Equation (1) agrees with previous derivations of bulk SCI [15,31]. The spin properties of the FM and NM channels needed for further analyses with equation (1) and equation (2) are presented in Supplemental Material S5 [23].

The spin absorption in our device can be due to spin relaxation within the W electrode, the W/WO$_x$ interface, the WO$_x$ layer, or the Cu/WO$_x$ interface. A careful analysis of the spin diffusion equation indicates that the spin absorption is dominated by the oxide layer (see Supplemental Material S6, [23]). The remaining question is now whether the absorption occurs in the "bulk" oxide layer or at the Cu/WO$_x$ interface. The absorption in the "bulk" oxide means that the spin current decays up to a certain depth $\leq t_{\text{WO}_x}$ within the oxide layer. We use equation (1) supposing that the oxide layer with resistivity $\rho_{\text{WO}_x}$ [Fig. 1(d)] is the SOM and the W just functions as a metallic electrode. $\lambda_{\text{WO}_x}$ is found to be $\sim 0.04$ nm (see Supplemental Material S6 [23]) which is smaller than the interatomic distance of typical transition metal oxides, excluding bulk absorption in the oxide layer.

Consequently, the spin absorption must take place at the Cu/WO$_x$ interface and equation (2) has to be used. Equation (2) is equivalent to equation (1) with the spin-loss conductance $G_\parallel A_\text{I}$ replacing the spin conductance $G_\text{s}^{\text{SOM}}$ (which depends on $\lambda_{\text{SOM}}$). Figure 2(d) presents $G_\parallel$ in our Py/Cu/W LSV obtained from $\Delta R_{\text{NL}}^{\text{abs}}/\Delta R_{\text{NL}}^{\text{ref}}$ at different temperatures [inset of Fig. 2(d)]. TABLE I compares the resulting $G_\parallel$ at the Cu/WO$_x$ interface to Cu/BiO$_x$ [10] and Cu/Au [12] interface, which are analyzed by the same method. $G_\parallel$ at the Cu/WO$_x$ interface ($\sim 20 \times 10^{13}$ $\Omega^{-1}\text{m}^{-2}$ at all temperatures) is remarkably larger than the ones observed in Cu/BiO$_x$ and Cu/Au.

TABLE I: Summary of the interfacial spin absorption and spin-charge interconversion in different interfaces at 10 K. The spin absorption is given by the interfacial spin-loss conductance $G_\parallel$ and the SCI by the interfacial spin-charge conductivity $\sigma_{SC}$. The inverse Edelstein length $\lambda_{IEE}$ is defined as $\sigma_{SC}/G_\parallel$.

| Materials system | $T$ [K] | $G_\parallel$ [$10^{13}$ $\Omega^{-1}\text{m}^{-2}$] | $\sigma_{\text{SC}}$ [$\Omega^{-1}\text{cm}^{-1}$] | $\lambda_{\text{IEE}}$ [nm] | Refs |
|---|---|---|---|---|---|
| Cu/BiO$_x$ | 10 | $2.8 \pm 0.2$ | $44 \pm 8$ | $0.16 \pm 0.03$ | [10] |
| Cu/Au | 10 | $7.6 \pm 0.6$ | $-127 \pm 8$ | $-0.17 \pm 0.04$ | [12] |
|  | 300 | $9.8 \pm 0.6$ | $-30 \pm 8$ | $-0.03 \pm 0.02$ | [12] |
| Cu/WO$_x$ | 10 | $21 \pm 2$ | $-1610 \pm 50$ | $-0.76 \pm 0.07$ | This work |
|  | 300 | $20 \pm 7$ | $-830 \pm 50$ | $-0.4 \pm 0.2$ | This work |

### C. Spin-charge interconversion

Next, we investigate if the spins absorbed at the Cu/WO$_x$ interface display interfacial SCI. Figure 3(a) shows the SEM image of the Py/Cu/W LSV, this time depicted together with the charge-to-spin measurement configuration. The bias current $I_{\text{bias}}$ is applied through the W electrode. As the charge current flows along the interface, and if the Cu/WO$_x$ interface contains any mechanism resulting in SCI, a spin accumulation will be created and diffuse into the Cu channel as a pure spin current. The magnetization of the Py electrode which is used to probe such spin current, by measuring a voltage between the Py and Cu electrodes ($V_{SC}$), is controlled by an external magnetic field ($H_x$) parallel to the polarization of the spin current.

Figure 3(b) presents the spin-charge resistance $R_{SC}$ ($= V_{SC}/I_{bias}$) as a function of $H_x$ for different temperatures between 10 K and 300 K. By sweeping $H_x$, a change in $R_{SC}$ is observed at all temperatures, indicating the occurrence of charge-to-spin conversion. This is unexpected considering the presence of the WO$_x$ layer. The difference between the low and high resistance, $2\Delta R_{SC}$, is called the spin-charge signal. The inset of Fig. 3(c) plots the temperature dependence of $2\Delta R_{SC}$ that is described (for the derivation see Supplemental Material S7 [23]) by:

$$\Delta R_{SC} = x_{WO_x} \frac{w_{NM}\sigma_{SC}}{t_{WO_x}\sigma_{WO_x}} \frac{G_s^{NM}}{G_s^{FM}G_{\parallel}A_I} \frac{4Pe^{L/2\lambda_{NM}}}{r_{FM}r_{\parallel}e^{L/\lambda_{NM}} + r_{\parallel} - 2}, \quad (3)$$

where $r_{\parallel} = 1 + 2G_s^{NM}/G_{\parallel}A_I$ and $\sigma_{SC}$ is the interfacial spin-charge conductivity which is the equivalent of the spin Hall conductivity for the bulk SCI. The electrical shunting factor $x_{WO_x}$ is essential for the evaluation of SCI in the LSV as it defines the charge current flowing along the interface. This is especially critical in our LSV, in which we have a structure of three different materials with a range of resistivities. The shunting factor of the Cu/WO$_x$ interface is assessed using 3D FEM simulations. Supplemental Material S8 [23] presents the details of electrical shunting in our device and the simulation to acquire the shunting factor.

We obtain $\sigma_{SC}$ by employing equation (3) and using the gathered values $\Delta R_{SC}$, $G_{\parallel}$ and $x_{WO_x}$. Figure 3(c) shows a $\sigma_{SC}$ value of $-1610$ $\Omega^{-1}$ cm$^{-1}$ at 10 K that decreases to $-830$ $\Omega^{-1}$ cm$^{-1}$ at 300 K. The negative sign of $\sigma_{SC}$ is also observed for the spin Hall conductivity $\sigma_{SH}$ in bulk $\beta$-W. Therefore, the sign seems sensible because, even though we could not extract the precise composition of the interfacial oxide layer, W is contained in the oxide layer. TABLE I indicates that $\sigma_{SC}$ of this interface is notably higher compared to the Cu/BiO$_x$ and Cu/Au systems. Remarkably, the observed $\sigma_{SC}$ is comparable to $\sigma_{SH}$ of Pt (~1600 $\Omega^{-1}$ cm$^{-1}$) [16].

Finally, we can combine the results of the interfacial spin-loss conductance ($G_{\parallel}$) and the interfacial spin-charge conductivity ($\sigma_{SC}$) to calculate the commonly used inverse Edelstein length $\lambda_{IEE}$ for the Cu/WO$_x$ interface. The resulting $\lambda_{IEE}$, $-0.76$ nm and $-0.4$ nm at 10 K and 300 K, respectively, are significantly larger than the ones observed in Cu/BiO$_x$ and Cu/Au at the same temperature (TABLE I) and $\theta_{SH}\lambda$ in Pt [16]. Note that it is more convenient to display the interfacial spin-loss and SCI parameters as $G_{\parallel}$ and $\sigma_{SC}$ because these are independent of the microscopic mechanisms leading to interfacial SCI (Rashba, spin-orbit filtering, etc.) and can be easily expanded to the 3D counterparts as $G_{\parallel}$ and $\sigma_{SC}$ are equivalent to the spin conductance of the SOM ($G_{SOM}$) and spin Hall conductivity ($\sigma_{SH}$) for bulk SCI, respectively. Furthermore, $\sigma_{SC}$ accounts for the Onsager reciprocity of the SCI at interfaces [10]. Figure 3(d) graphs $\lambda_{IEE} = \sigma_{SC}/G_{\parallel}$ as a function of temperature, showing a decrease of this parameter with increasing temperature similar to the behavior observed in the Cu/BiO$_x$ and Cu/Au systems.

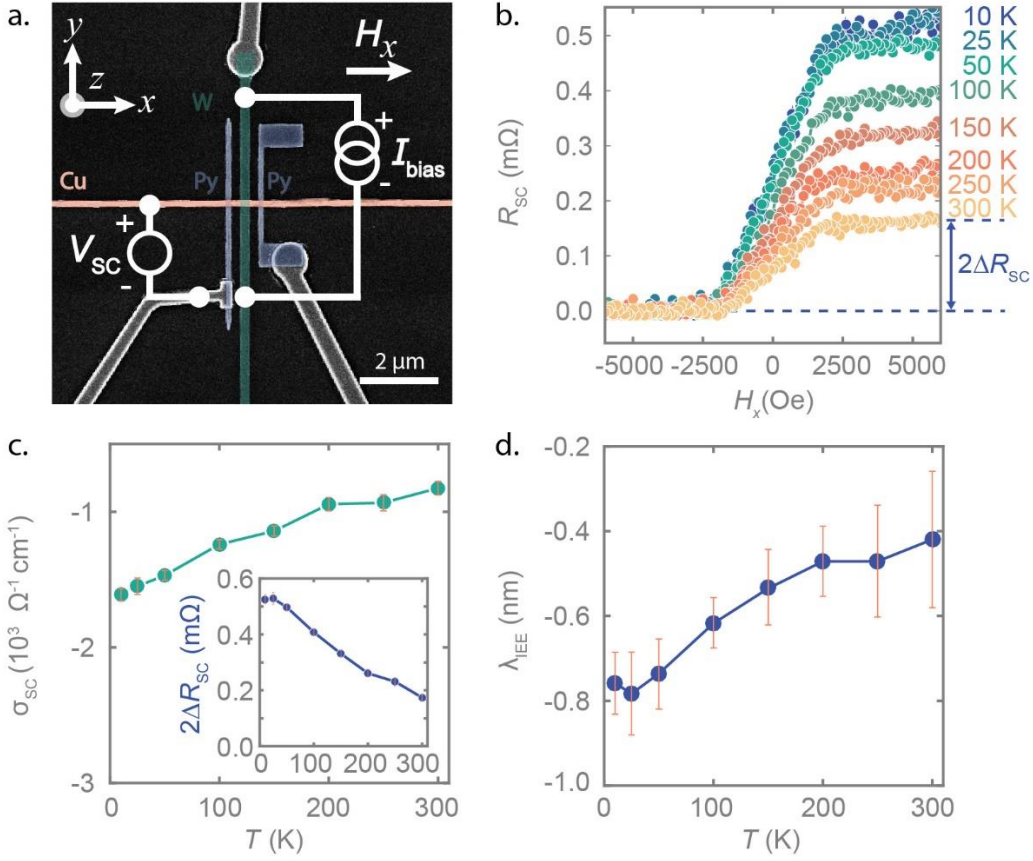

FIG. 3: Charge-to-spin conversion at the Cu/WOₓ interface. (a) A false-color SEM image of the Py/Cu/W LSV and the charge-to-spin conversion measurement configuration. The external magnetic field is oriented in the x-direction. (b) The magnetic field dependence of the spin-charge resistance $R_{SC}$ for various temperatures. $R_{SC}$ is the average of the magnetic field trace and retrace. An offset is added to $R_{SC}$ such that the low resistance state is zero. The difference between the low and high resistance state is the spin-charge signal $2\Delta R_{SC}$. (c) The temperature dependence of the interfacial spin-charge conductivity $\sigma_{SC}$ for the Cu/WOₓ calculated with equation (3) combined with the obtained values for $\Delta R_{SC}$, $G_\parallel$, and $x_{WO_x}$. Inset: $R_{SC}$ as a function of temperature. (d) The temperature dependence of inverse Edelstein length $\lambda_{IEE}$ resulting from the ratio between $\sigma_{SC}$ and $G_\parallel$ [Fig. 2(d)].

Although the device was not fabricated to study the interfacial SCI of a Cu/WOₓ interface, spin-orbit torque studies on W/CoFe [32] and CoFe/WOₓ structures [33] show large SCI efficiencies that are discussed to be of interfacial origin. Additionally, enhanced SCI efficiencies are observed in CuW alloys [34] and highly efficient SCI is measured in Cu/AlOₓ interfaces [35] using spin-torque ferromagnetic resonance. The lack of heavy elements in this Cu/AlOₓ structure reveals SCI mediated through orbital transport. This type of orbital Hall effect and orbital Edelstein effect could also occur in our Cu/WOₓ structure. Either way, the large $\lambda_{IEE}$ makes this system potentially interesting for the magnetic-state readout in the MESO device [21,22].

## IV. CONCLUSIONS

We studied a Py/Cu/W LSV to investigate the spin-charge interconversion properties of W. A careful analysis leads to the discovery of an oxide layer at the Cu/W interface with a high resistivity that prevents spin absorption into the W electrode. We observe that the spin absorption happens at the Cu/WOₓ interface with a temperature-independent interfacial spin-loss conductance of $G_\parallel \approx 20 \times 10^{13}\,\Omega^{-1}\,m^{-2}$. Additionally, a large interfacial SCI is observed, with $\sigma_{SC}$ varying from $-1610\,\Omega^{-1}\,cm^{-1}$ to − 830 $\Omega^{-1}$ cm$^{-1}$ for the temperature range 10 K to 300 K. The system has an efficiency given by the inverse Edelstein length $\lambda_{\text{IEE}}$ that ranges from −0.76 nm to −0.4 nm when varying from 10 K to 300 K, which is substantially larger than other metallic interfaces [10,12] or even Pt [16]. The large SCI efficiency in the Cu/WO$_x$ interface could be applicable to the proposed MESO logic device [21,22]. Importantly, our study indicates that one must characterize interfaces carefully when studying the spin transport properties of spin-orbit materials in nanodevices.

## ACKNOWLEDGEMENTS

This work is supported by Intel Corporation through the Semiconductor Research Corporation under MSR-INTEL TASK 2017-IN-2744 and the 'FEINMAN' Intel Science Technology Center, and by the Spanish MICINN under the Maria de Maeztu Units of Excellence Programme (MDM-2016-0618 and CEX2020-001038-M) and under project Nos. RTI2018-094861-B-I00, PID2020-114252GB-I00, PID2020-112811GB-I00, and PID2021-122511OB-I00. W.Y.C. acknowledges postdoctoral fellowship support from "Juan de la Cierva Formación" programme by the Spanish MICINN (grant No. FJC 2018-038580-I). E.S. thanks the Spanish MECD for a Ph.D. fellowship (grant No. FPU14/03102). D.C.V. acknowledges support from the European Commission for a Maria Sklodowska-Curie individual fellowship (grant No. 892983-SPECTER). N.O. thanks the Spanish MICINN for a PhD fellowship (grant No. BES-2017-07963). The work of S.I. and F.S.B. is supported by European Union's Horizon 2020 Research and Innovation Framework Programme under Grant No. 800923 (SUPERTED). I.V.T. and F.S.B. acknowledge support by Grupos Consolidados UPV/EHU del Gobierno Vasco (Grant Nos. IT1249-19 and IT1591-22). F.S.B. acknowledges financial support by the A. v. Humboldt Foundation.

## REFERENCES

[1] A. Manchon, J. Železný, I. M. Miron, T. Jungwirth, J. Sinova, A. Thiaville, K. Garello, and P. Gambardella, *Current-Induced Spin-Orbit Torques in Ferromagnetic and Antiferromagnetic Systems*, Rev. Mod. Phys. **91**, 035004 (2019).

[2] P. Barla, V. K. Joshi, and S. Bhat, *Spintronic Devices: A Promising Alternative to CMOS Devices*, Vol. 20 (Springer US, 2021).

[3] S. Manipatruni, D. E. Nikonov, and I. A. Young, *Beyond CMOS Computing with Spin and Polarization*, Nat. Phys. **14**, 338 (2018).

[4] J. Sinova, S. O. Valenzuela, J. Wunderlich, C. H. Back, and T. Jungwirth, *Spin Hall Effects*, Rev. Mod. Phys. **87**, 1213 (2015).

[5] J. C. R. Sánchez, L. Vila, G. Desfonds, S. Gambarelli, J. P. Attané, J. M. De Teresa, C. Magén, and A. Fert, *Spin-to-Charge Conversion Using Rashba Coupling at the Interface between Non-Magnetic Materials*, Nat. Commun. **4**, 2944 (2013).

[6] S. Karube, K. Kondou, and Y. C. Otani, *Experimental Observation of Spin-to-Charge Current Conversion at Non-Magnetic Metal/Bi$_2$O$_3$ Interfaces*, Appl. Phys. Express **9**, 033001 (2016).

[7] J. C. Rojas-Sánchez et al., *Spin to Charge Conversion at Room Temperature by Spin Pumping into a New Type of Topological Insulator: α -Sn Films*, Phys. Rev. Lett. **116**, 096602 (2016).

[8] H. Tsai, S. Karube, K. Kondou, N. Yamaguchi, F. Ishii, and Y. Otani, *Clear Variation of Spin Splitting by Changing Electron Distribution at Non-Magnetic Metal/Bi2O3 Interfaces*, Sci. Rep. **8**, 5564 (2018).

[9]  D. C. Vaz et al., *Mapping Spin-Charge Conversion to the Band Structure in a Topological Oxide Two-Dimensional Electron Gas*, Nat. Mater. **18**, 1187 (2019).

[10] C. Sanz-Fernández, V. T. Pham, E. Sagasta, L. E. Hueso, I. V. Tokatly, F. Casanova, and F. S. Bergeret, *Quantification of Interfacial Spin-Charge Conversion in Hybrid Devices with a Metal/Insulator Interface*, Appl. Phys. Lett. **117**, 142405 (2020).

[11] L. M. Vicente-Arche et al., *Spin–Charge Interconversion in KTaO3 2D Electron Gases*, Adv. Mater. **33**, 2102102 (2021).

[12] V. T. Pham, H. Yang, W. Y. Choi, A. Marty, I. Groen, A. Chuvilin, F. S. Bergeret, L. E. Hueso, I. V. Tokatly, and F. Casanova, *Large Spin-Charge Interconversion Induced by Interfacial Spin-Orbit Coupling in a Highly Conducting All-Metallic System*, Phys. Rev. B **104**, 184410 (2021).

[13] G. Allen, S. Manipatruni, D. E. Nikonov, M. Doczy, and I. A. Young, *Experimental Demonstration of the Coexistence of Spin Hall and Rashba Effects in β-Tantalum/Ferromagnet Bilayers*, Phys. Rev. B **91**, 144412 (2015).

[14] Y. Du, H. Gamou, S. Takahashi, S. Karube, M. Kohda, and J. Nitta, *Disentanglement of Spin-Orbit Torques in Pt/Co Bilayers with the Presence of Spin Hall Effect and Rashba-Edelstein Effect*, Phys. Rev. Appl. **13**, 054014 (2020).

[15] M. Morota, Y. Niimi, K. Ohnishi, D. H. Wei, T. Tanaka, H. Kontani, T. Kimura, and Y. Otani, *Indication of Intrinsic Spin Hall Effect in 4d and 5d Transition Metals*, Phys. Rev. B **83**, 174405 (2011).

[16] E. Sagasta, Y. Omori, M. Isasa, M. Gradhand, L. E. Hueso, Y. Niimi, Y. C. Otani, and F. Casanova, *Tuning the Spin Hall Effect of Pt from the Moderately Dirty to the Superclean Regime*, Phys. Rev. B. **94**, 060412(R) (2016).

[17] J. C. Rojas-Sánchez, N. Reyren, P. Laczkowski, W. Savero, J. P. Attané, C. Deranlot, M. Jamet, J. M. George, L. Vila, and H. Jaffrès, *Spin Pumping and Inverse Spin Hall Effect in Platinum: The Essential Role of Spin-Memory Loss at Metallic Interfaces*, Phys. Rev. Lett. **112**, 106602 (2014).

[18] W. Zhang, W. Han, X. Jiang, S. H. Yang, and S. S. P. Parkin, *Role of Transparency of Platinum-Ferromagnet Interfaces in Determining the Intrinsic Magnitude of the Spin Hall Effect*, Nat. Phys. **11**, 496 (2015).

[19] C. F. Pai, Y. Ou, L. H. Vilela-Leão, D. C. Ralph, and R. A. Buhrman, *Dependence of the Efficiency of Spin Hall Torque on the Transparency of Pt / Ferromagnetic Layer Interfaces*, Phys. Rev. B **92**, 064426 (2015).

[20] V. T. Pham, M. Cosset-Chéneau, A. Brenac, O. Boulle, A. Marty, J.P. Attané, and L. Vila, *Evidence of Interfacial Asymmetric Spin Scattering at Ferromagnet-Pt Interfaces*, Phys. Rev. B **103**, L201403 (2021).

[21] S. Manipatruni, D. E. Nikonov, C. Lin, T. A. Gosavi, H. Liu, B. Prasad, Y. Huang, E. Bonturim, R. Ramesh, and I. A. Young, *Scalable Energy-Efficient Magnetoelectric Spin–Orbit Logic*, Nature **565**, 35 (2019).

[22] V. T. Pham et al., *Spin–Orbit Magnetic State Readout in Scaled Ferromagnetic/Heavy Metal Nanostructures*, Nat. Electron. **3**, 309 (2020).

[23] *Supplemental Material: Enhanced Interfacial Spin-Charge Interconversion at an Oxidized Cu/W Interface*, (n.d.).

[24] Q. Hao, W. Chen, and G. Xiao, *Beta (β) Tungsten Thin Films: Structure, Electron Transport, and Giant Spin Hall Effect*, Appl. Phys. Lett. **106**, 182403 (2015).

[25] R. J. Pedersen and F. L. Vernon, *Effect of Film Resistance on Low-Impedance Tunneling Measurements*, Appl. Phys. Lett. **10**, 29 (1967).

[26] J. M. Pomeroy and H. Grube, *"Negative Resistance" Errors in Four-Point Measurements of Tunnel Junctions and Other Crossed-Wire Devices*, J. Appl. Phys. **105**, 094503 (2009).

[27] E. Sagasta, Y. Omori, S. Vélez, R. Llopis, C. Tollan, A. Chuvilin, L. E. Hueso, M. Gradhand, Y. C. Otani, and F. Casanova, *Unveiling the Mechanisms of the Spin Hall Effect in Ta*, Phys. Rev. B. **98**, 060410(R) (2018).

[28] Y. Niimi, M. Morota, D. H. Wei, C. Deranlot, M. Basletic, A. Hamzic, A. Fert, and Y. Otani, *Extrinsic Spin Hall Effect Induced by Iridium Impurities in Copper*, Phys. Rev. Lett. **106**, 126601 (2011).

[29] Y. Niimi, Y. Kawanishi, D. H. Wei, C. Deranlot, H. X. Yang, M. Chshiev, T. Valet, A. Fert, and Y. Otani, *Giant Spin Hall Effect Induced by Skew Scattering from Bismuth Impurities inside Thin Film CuBi Alloys*, Phys. Rev. Lett. **109**, 156602 (2012).

[30] P. Laczkowski et al., *Large Enhancement of the Spin Hall Effect in Au by Side-Jump Scattering on Ta Impurities*, Phys. Rev. B **96**, 140405(R) (2017).

[31] W. Yan, E. Sagasta, M. Ribeiro, Y. Niimi, L. E. Hueso, and F. Casanova, *Large Room Temperature Spin-to-Charge Conversion Signals in a Few-Layer Graphene/Pt Lateral Heterostructure*, Nat. Commun. **8**, 661 (2017).

[32] Y. Takeuchi, C. Zhang, A. Okada, H. Sato, S. Fukami, and H. Ohno, *Spin-Orbit Torques in High-Resistivity-W/CoFeB/MgO*, Appl. Phys. Lett. **112**, 192408 (2018).

[33] K.-U. Demasius, T. Phung, W. Zhang, B. P. Hughes, S.-H. Yang, A. Kellock, W. Han, A. Pushp, and S. S. P. Parkin, *Enhanced Spin–Orbit Torques by Oxygen Incorporation in Tungsten Films*, Nat. Commun. **7**, 10644 (2016).

[34] B. Coester, G. D. H. Wong, Z. Xu, J. Tang, W. L. Gan, and W. S. Lew, *Enhanced Spin Hall Conductivity in Tungsten-Copper Alloys*, J. Magn. Magn. Mater. **523**, 167545 (2021).

[35] J. Kim, D. Go, H. Tsai, D. Jo, K. Kondou, H. W. Lee, and Y. C. Otani, *Nontrivial Torque Generation by Orbital Angular Momentum Injection in Ferromagnetic-Metal*, Phys. Rev. B **103**, L020407 (2021).

[36] E. Sagasta et al., *Spin Diffusion Length of Permalloy Using Spin Absorption in Lateral Spin Valves Spin Diffusion Length of Permalloy Using Spin Absorption in Lateral Spin Valves*, Appl. Phys. Lett. **111**, 082407 (2017).

[37] T. Kimura, Y. Otani, T. Sato, S. Takahashi, and S. Maekawa, *Room-Temperature Reversible Spin Hall Effect*, Phys. Rev. Lett. **98**, 156601 (2007).

[38] Y. Omori, F. Auvray, T. Wakamura, Y. Niimi, A. Fert, and Y. Otani, *Inverse Spin Hall Effect in a Closed Loop Circuit*, Appl. Phys. Lett. **104**, 242415 (2014).

# Supplemental Material

## S1. Resistivities of the different electrodes in the Py/Cu/W lateral spin valve

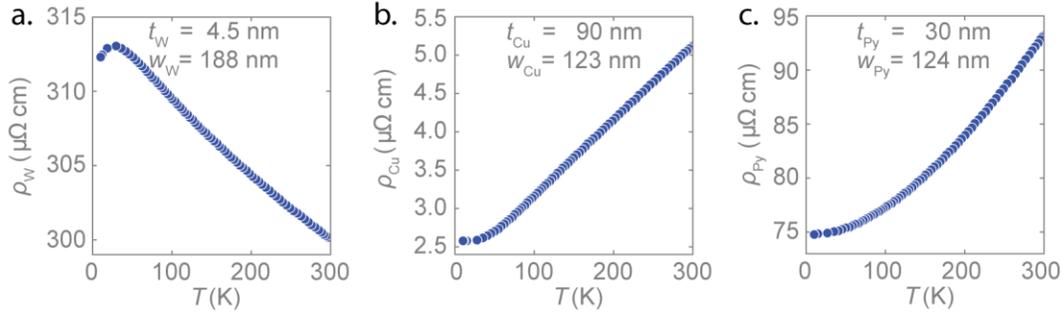

FIG. S1: The resistivities of electrodes in the Py/Cu/W LSV. The temperature dependence of the resistivity of the (a) W electrode; (b) Cu channel; and (c) Py electrode.

## S2. STEM characterization of the Cu/W interface

The cross-section of the device [Fig. 1(a)] has been prepared by a standard focused ion beam (FIB) method and fixed on the Cu half-grid. STEM/EDX study was performed on Titan 60-300 TEM/STEM microscope (FEI, Netherlands) at 300 kV accelerating voltage. The microscope was equipped with RTEM EDX detector (EDAX, UK). Along with high-angle annular dark field (HAADF) STEM images, EDX spectrum data cubes were acquired from selected regions. 3D spectral data were processed in Digital Micrograph software (Gatan, UK) using embedded Multiple Linear Least Square (MLLS) fitting procedure.

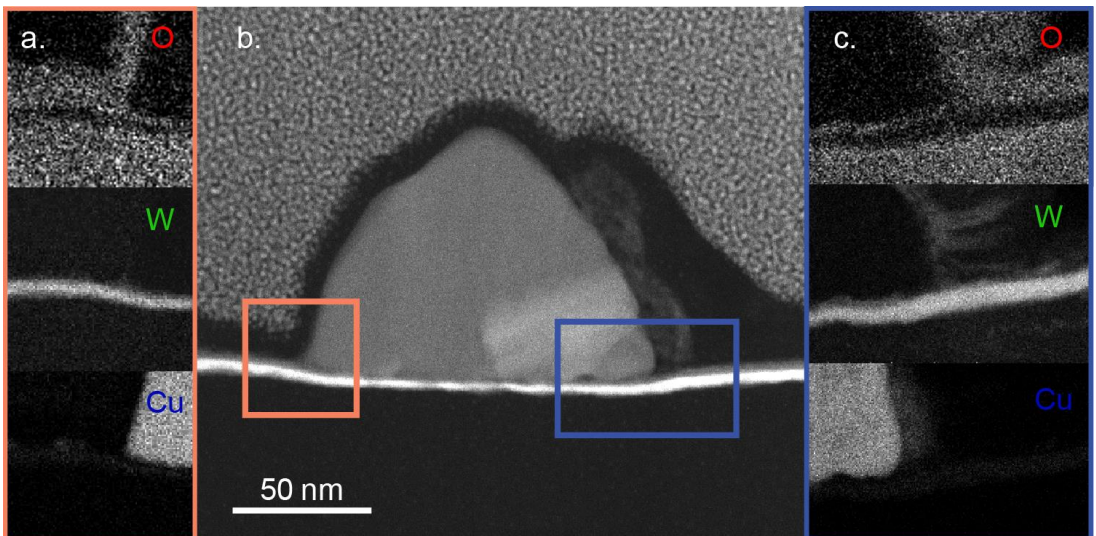

FIG. S2.1: STEM image of the Cu/W interface. (b) The large overview image shows the LSV cross-section of the transversely cut Cu channel (grey) and longitudinally cut W electrode (white). (a) The blue square and (c) orange square areas where EDX analyses have been performed. The element analysis identifies O, W, and Cu. Note that the images reveal the presence of O at the Cu/W interface.

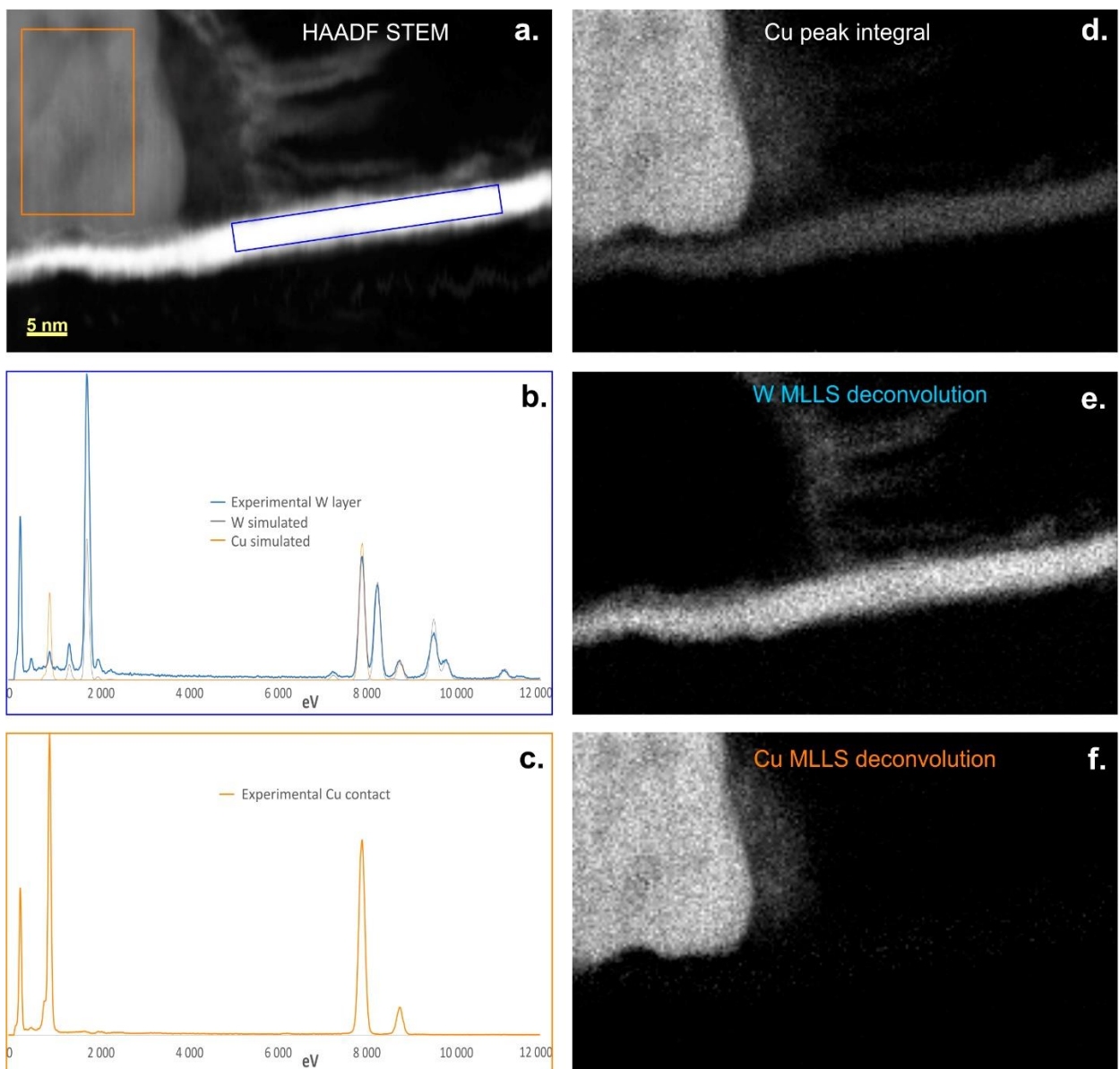

*Fig S2.2: Illustration of the EDX spectra deconvolution procedure. (a) HAADF STEM image of analyzed area. Note that this area is also partly shown in Fig. 1(c) of the main text. Blue rectangle marks the area for spectrum on panel (b), orange rectangle marks the area for spectrum on panel (c); (b) Spectrum from blue rectangle on panel (a). Simulated spectra for Cu and W are overlayed. Though the layer is solely W, a strong Cu signal is also observed as an artifact. (c) Spectrum of a Cu contact [orange rectangle on panel (a)]; (d) Cu distribution map obtained as an integral of Cu peaks intensity; (e) Fit coefficient for the blue spectrum on panel (b) in MLLS fitting; (f) Fit coefficient for the spectrum on panel (c) in MLLS fitting.*

Figure S2.1(b) shows the cross-sectional STEM image of the Cu/W interface in the Py/Cu/W LSV. EDX is used to map the composition of the regions indicated by the orange and blue rectangles. The blue rectangle is the same area as presented in Fig. 1(a) of the main text. Figures S2.1(a) and (c) display the observed elements (O, W and Cu) in these orange and blue regions, respectively. Detailed information about the W electrode can be acquired from the STEM images, such as the thickness of the W below the Cu channel ($t_W \sim 2.8$ nm). This is different from the electrode thickness of the W ($t_W^* \sim 4.5$ nm) because of the Ar-ion milling during the fabrication. The natural $WO_x$ on top of the W electrode (outside the interface area with Cu) is about 3 nm.

The EDX also shows the presence of O at the interface between the Cu and W electrodes. The oxidized layer has a thickness of $\sim 1.5$ nm. During the fabrication of the LSV, Ar-ion milling for $WO_x$ removal combined with *in-situ* sputtering of a thin layer of Cu, should have protected the interface from oxidization. The natural $WO_x$ below the Cu has been removed during the Ar-ion milling as $t_W < t_W^*$. The oxidation has, therefore, taken place during the fast transfer from the ion-miller to the evaporator ($\sim 3$ min). This can be due to several reasons, for example, the *in-situ* sputtered Cu layer is not thick enough or does not grow homogeneously on the W electrode and, therefore, does not fully protect the W from oxidation.

In the EDX analysis of Cu distribution, we observe a typical artifact reflecting in Cu signal coming from any dense part of the sample independently of the composition. Figure S2.2(b) shows a spectrum from the area marked by blue rectangle in Fig. S2.1(b). Though by preparation the bright layer contains only W, there are strong Cu peaks visible in the spectrum as well as Cu distribution map obtained by integrating Cu peaks [Fig. S2.2(d)] shows the presence of Cu in the W layer. This artifact appears due to characteristic X-ray generation by secondary electrons landing on the Cu grid supporting the lamellae. To decouple Cu and W signals, we have used the spectrum from W layer [Fig. S2.2(b)] as a characteristic spectrum of W, the spectrum from Cu contact [Fig. S2.2(c)] as a signature for Cu and have performed a MLLS fitting (simulated spectrum was used for O). The outcome of MLLS deconvolution is shown on Figs. S2.2(e) and S2.2(f). Cu signal is efficiently eliminated from the W layer, which in particular allows unambiguous interpretation of the interface oxide as $WO_x$ [See Fig. 1(c) of the main text].

## S3. 3D FEM simulation for the $WO_x$ resistivity

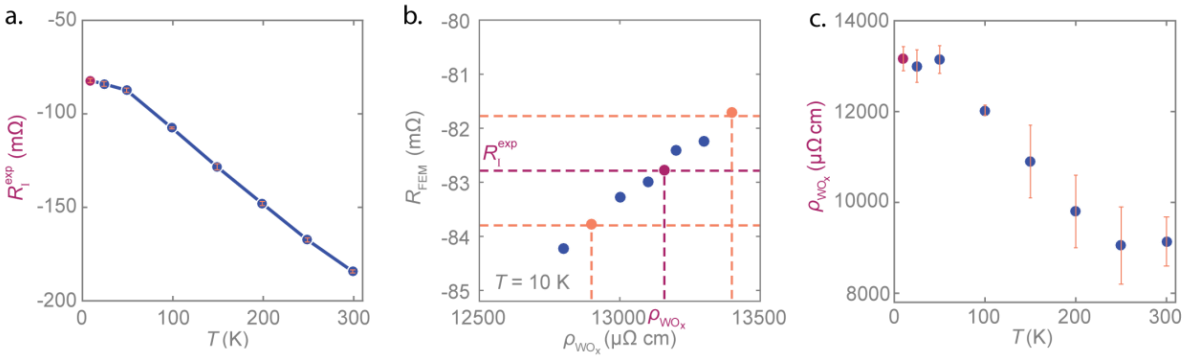

*FIG. S3: Analyses of the 3D FEM simulation to obtain the $WO_x$ resistivity. (a) The temperature dependence of the experimentally measured interface resistance. The value at 10 K is indicated in purple and will be evaluated in panel b. (b) The FEM interface resistance extracted from the simulation for different resistivities of the $WO_x$ layer at 10 K. The experimental data point and its associated error are displayed in purple and orange, respectively. The actual resistivities of the $WO_x$ layer is given by $R_{FEM} = R_I^{exp}$. (c) The temperature dependence of the $WO_x$ resistivity.*

## S4. Derivation of the spin signal and the spin absorption in a lateral spin valve

In the LSV, a non-local technique can be used where injection and detection of pure spin current are separated by a spin transport channel as shown in Fig. S4(a). A charge current is applied from the NM into a FM, creating a spin accumulation at the FM/NM interface. The spin accumulation will diffuse away from the FM within the NM (+x-direction) as a pure spin current with a characteristic spin diffusion length $\lambda_{\text{NM}}$. The detector, a second FM electrode, is placed at a distance $L$ (with $L \sim \lambda_{\text{NM}}$) from the injector. Additionally, an electrode is positioned in between the injector and detector. Figure S4(b) illustrates a LSV consisting of two FM electrodes (FM1 and FM2) that are connected via a NM channel with a long spin diffusion length and a middle electrode of SOM. If the middle electrode absorbs part of the spin current, the spin properties of this electrode can be obtained. Figure S4(c) points out the dimensions and the coordinates of the LSV which are used in the upcoming derivation.

The spin flow in the device for spin absorption [Fig. S4(b)] can be described by the one-dimensional (1D) spin-diffusion model. Within this model, the spin currents are assumed to only have components in $z$ and $x$ directions. Therefore, the spin diffusion equation in the NM, FMs and SOM electrodes are, respectively:

$$\frac{\partial^2 \mu_s^{\text{NM}}}{\partial x^2} = \frac{\mu_s^{\text{NM}}}{\lambda_{\text{NM}}^2}, \tag{S1}$$

$$\frac{\partial^2 \mu_s^{\text{FM}}}{\partial z^2} = \frac{\mu_s^{\text{FM}}}{\lambda_{\text{FM}}^2}, \tag{S2}$$

$$\frac{\partial^2 \mu_s^{\text{SOM}}}{\partial z^2} = \frac{\mu_s^{\text{SOM}}}{\lambda_{\text{SOM}}^2}. \tag{S3}$$

In the 1D approximation, the spin properties of the NM electrode in the volume below the intersection with the SOM electrode correspond to a single point x = 0, that being the case, the boundary conditions at x = 0 are

$$\mu_s^{\text{NM}}|_{0^+} = \mu_s^{\text{NM}}|_{0^-}, \tag{S4}$$

$$A_{\text{NM}} \left( j_{s,x}^{\text{NM}}|_{0^-} - j_{s,x}^{\text{NM}}|_{0^+} \right) = A_{\text{I}} j_{s,z}^{\text{NM}}|_0, \tag{S5}$$

where $A_{\text{NM}} = t_{\text{NM}} w_{\text{NM}}$ is the cross-sectional area of the NM electrode and $A_{\text{I}} = w_{\text{SOM}} w_{\text{NM}}$ is the NM/SOM interface area. At the NM/SOM interface ($z = 0$), the boundary conditions read

$$\left( j_{s,z}^{\text{SOM}} - J_{s,z}^{\text{NM}} \right)\big|_{z=0} = -G_\| A_{\text{I}} \frac{\mu_s^{\text{NM}} + \mu_s^{\text{SOM}}}{2}\bigg|_{x=0, z=0}, \tag{S6}$$

$$\frac{1}{2} \left( j_{s,z}^{\text{SOM}} + j_{s,z}^{\text{NM}} \right)\big|_{z=0} = G_{\text{I}} (\mu_s^{\text{NM}} - \mu_s^{\text{SOM}})\big|_{x=0, z=0}, \tag{S7}$$

where $G_\|$ is the interfacial spin-loss conductance and $G_{\text{I}}$ is the interface conductance of the NM/SOM interface. We assume that $w_{\text{SOM}} \ll \lambda_{\text{NM}}$ such that $\mu_s^{\text{NM}}$ below the interface can be considered constant.

The amount of spin injected and detected also depends on the boundary conditions at the NM/FM interfaces given by the continuity of $j_s$:

$$A_{\text{NM}} \left( j_{s,x}^{\text{NM}}\big|_{-\frac{L}{2}^-} - j_{s,x}^{\text{NM}}\big|_{-\frac{L}{2}^+} \right) = A_{\text{I}}^{\text{FM1}} \left( j_{s,z}^{\text{FM1}} - eP_{\text{FM1}} j_{\text{bias}} \right)\big|_{z=0}, \tag{S8}$$

$$A_{\text{NM}} \left( j_{s,x}^{\text{NM}}\big|_{\frac{L}{2}^-} - j_{s,x}^{\text{NM}}\big|_{\frac{L}{2}^+} \right) = A_{\text{I}}^{\text{FM2}} j_{s,z}^{\text{FM2}}\big|_{z=0}. \tag{S9}$$

where $A_\text{I}^\text{FM}$ is area of the NM/FM interfaces. The $P_\text{FM1}$ is the polarization of FM1 and $j_\text{bias}$ is the charge current density injected at $x = -L/2$. The NM/FM interfaces are considered transparent, such that continuity of the chemical potentials at the interface give:

$$\left(\mu_\text{s}^\text{NM} - \mu_\text{s}^\text{FM1}\right)\Big|_{x=-\frac{L}{2},\ z=0} = 0, \tag{S10}$$

$$\left(\mu_\text{s}^\text{NM} - \mu_\text{s}^\text{FM2}\right)\Big|_{x=\frac{L}{2},\ z=0} = 0. \tag{S11}$$

The spin accumulation within the NM electrode decays exponentially as:

$$\mu_\text{s}^\text{NM}(x) = A e^{-|x|/\lambda_\text{NM}} + B e^{-\left|x+\frac{L}{2}\right|/\lambda_\text{NM}} + C e^{-\left|x-\frac{L}{2}\right|/\lambda_\text{NM}}, \tag{S12}$$

and the spin current density is:

$$j_{s,x}^\text{NM} = \frac{\sigma_\text{NM}}{\lambda_\text{NM}}\left[ A\,\text{sign}(x) e^{-\frac{|x|}{\lambda_\text{NM}}} + B\,\text{sign}\left(x+\frac{L}{2}\right)e^{-\frac{\left|x+\frac{L}{2}\right|}{\lambda_\text{NM}}} + C\,\text{sign}\left(x-\frac{L}{2}\right)e^{-\frac{\left|x-\frac{L}{2}\right|}{\lambda_\text{NM}}} \right].$$

The solution of the spin diffusion equation in the FM electrodes become:

$$\mu_\text{s}^\text{FM}(z) = D e^{-z/\lambda_\text{FM}}, \tag{S13}$$

$$j_{s,z}^\text{FM} = -\frac{\sigma_\text{FM}}{\lambda_\text{FM}}(1 - P^2) D e^{-z/\lambda_\text{FM}}, \tag{S14}$$

and in the SOM electrode they are:

$$\mu_\text{s}^\text{SOM}(z) = E e^{-z/\lambda_\text{SOM}}, \tag{S15}$$

$$j_{s,z}^\text{SOM} = -\frac{\sigma_\text{SOM}}{\lambda_\text{SOM}} E e^{-z/\lambda_\text{SOM}}. \tag{S16}$$

A, B, C, D and E are integration constants. FM1 and FM2 are respectively located at $x = -L/2$ and $x = +L/2$ and SOM is placed at $x = 0$.

The detected non-local voltage is

$$eV_\text{NL} = P_\text{FM2}\mu_\text{s}^\text{FM2}. \tag{S17}$$

and the corresponding non-local resistance is

$$R_\text{NL} = \frac{V_\text{NL}}{j_\text{bias} A_\text{I}^\text{FM1}}. \tag{S18}$$

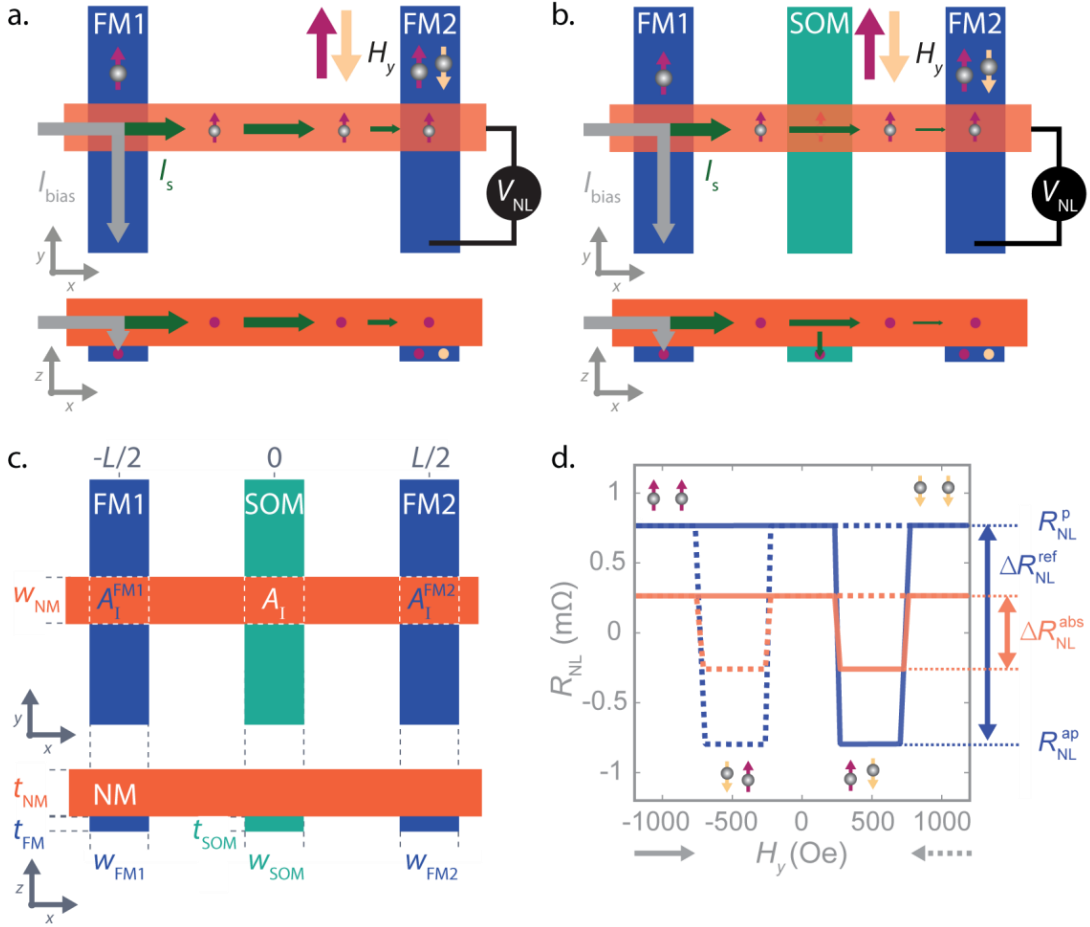

FIG. S4: (a) A reference LSV has a FM injector and detector bridged by a NM channel. A charge current $I_{bias}$ (grey arrows) is applied between the injector and the NM to inject a spin current (green arrows). The detector will measure a different potential $V_{NL}$ depending on the magnetization of the FM with respect to the spin current. The FM electrodes are designed such that the magnetization switches at different magnetic field. This permits observation of the parallel and antiparallel configuration when sweeping an external magnetic field ($H_y$). (b) The LSV for spin absorption has a FM injector and detector and a SOM middle electrode. The SOM will absorb part of the spin current that is flowing in the NM channel (see green arrows) reducing the spin current that arrives to the detector. (c) Important device dimensions for the derivations of the spin flow and SCI in the LSV. The width and thickness are given by $w_i$ and $t_i$ for $i$ = NM, FM1, FM2 and SOM and the interface area $A_I^j$ with $j$ = FM1, FM2 and SOM. The FM interelectrode distance is L where FM1, SOM and FM2 are positioned at $x = -L/2, x = 0, x = L/2$, respectively. The NM/SOM interface is at $z = 0$. (d) The magnetic field dependence of the non-local spin resistance $R_{NL} = V_{NL}/I_{bias}$ for the reference LSV ($R_{NL}^{ref}$, blue line) and the absorption LSV ($R_{NL}^{abs}$, orange line). The solid and dashed lines show the trace and retrace. The non-local spin signal $\Delta R_{NL}$, the difference in resistance between the parallel and antiparallel state $R_{NL}^p - R_{NL}^{ap}$, shows $\Delta R_{NL}^{abs} < \Delta R_{NL}^{ref}$ indicating that the SOM electrode absorbs some of the spin current. The amount of spin current absorbed can be related to the spin absorption properties of the SOM.

Figure S4(d) shows $R_{NL}$ while sweeping an external magnetic field $H_y$. We assume that FM1 and FM2 are equivalent ($P_{FM1} = P_{FM2} = P, w_{FM1} = w_{FM2}$ and $t_{FM1} = t_{FM2}$) but designed in such a way that the switching field, the magnetic field at which the magnetization $m$ reverses, are different for each FM electrode. Saturating $m$ such that the FMs are parallel (↑↑) gives a positive $R_{NL}$. While sweeping $H_y$ through zero, $m$ of one of the FM reverses first due to the difference in switching fields, resulting in an antiparallel magnetic configuration (↑↓). When $H_y$ continues to increase, the second FM will switch, and the magnetic configuration is parallel again (↓↓). Even though this parallel configuration is opposite to the starting configuration, $R_{NL}$ is the same because the chemical potential difference between the two electrodes is probed. The difference in $R_{NL}$ between the parallel and antiparallel configuration is the non-local spin signal given by $\Delta R_{NL} = R_{NL}^p - R_{NL}^{ap}$ which allows removing any baseline in the signal. When considering all the above, the spin absorbed signal is defined as:

$$\Delta R_{NL}^{abs} = \left[8R_s^{NM}P^2Q_{FM}R_s^{FM}e^{L/\lambda_{NM}}[2Q_\parallel(r-1)+Q_{RI}Q_{SOM}]\right]/[r_{FM}^2 e^{2L/\lambda_{NM}}[4Q_\parallel r + 2(Q_{RI}+2)Q_{SOM} + Q_{RI}] - 2r_{FM}e^{L/\lambda_{NM}}(4Q_\parallel + Q_{RI} + 4Q_{SOM}) + 4Q_\parallel(2-r) + 2(2-Q_{RI})Q_{SOM} + Q_{RI}]. \quad (S20)$$

$r_{FM} = 1 + 2G_s^{NM}/G_s^{FM}$ considering a transparent FM/NM interface and $r = 1 + 2G_s^{NM}/G_s^{SOM} + 2G_s^{NM}/G_I$ introducing the spin conductances: $G_s^{NM} = w_{NM}t_{NM}/\rho_{NM}\lambda_{NM}$, $G_s^{FM} = (1-P^2)w_{NM}w_{FM}/\rho_{FM}\lambda_{FM}$, and $G_s^{SOM} = w_{NM}w_{SOM}/\rho_{SOM}\lambda_{SOM}$. Furthermore, $Q_{FM} = G_s^{NM}/G_s^{FM}$ and $Q_{SOM} = G_s^{NM}/G_s^{SOM}$. The NM/SOM interface is represented by $Q_{RI} = G_s^{NM}/G_I$ and $Q_\parallel = G_s^{NM}/G_\parallel A_I$.

If we remove the middle SOM electrode, we obtain the spin reference signal:

$$\Delta R_{NL}^{ref} = 4R_s^{NM}P^2 \frac{e^{\frac{L}{\lambda_{NM}}}}{r_{FM}^2 e^{2L/\lambda_{NM}}-1}. \quad (S21)$$

Finally, we can calculate the ratio of $\Delta R_{NL}^{abs}/\Delta R_{NL}^{ref} = \eta$, that is,

$$\eta = \frac{2(r_{FM}e^{\frac{L}{\lambda_{NM}}}+1)[2Q_\parallel(r-1)+Q_{RI}Q_{SOM}]}{r_{FM}e^{\frac{L}{\lambda_{NM}}}(4Q_\parallel r + 2(Q_{RI}+2)Q_{SOM}+4Q_{RI})+4Q_\parallel(r-2)-2(2-2Q_{RI})Q_{SOM}+Q_{RI}}. \quad (S22)$$

Equation (S22) includes both the spin absorption of the bulk ($Q_{SOM}$) and the interface ($Q_\parallel$ and $Q_{RI}$). However, when the bulk and interface properties are on the same order of magnitude, there are two unknown parameters for the same contribution and no weight can be given to them such that bulk and interface contributions cannot be distinguished. But, if one of the two dominate over the other, the spin properties can be extracted considering the following two limiting cases:

Bulk contribution including an interface conductance ($G_I \neq 0$) but no interfacial spin-loss:

$$\eta_{bulk} = \frac{\Delta R_{NL}^{abs}}{\Delta R_{NL}^{ref}}\bigg|_{\substack{G_\parallel=0 \\ Q_\parallel=\infty}} = \left[1 + \frac{1}{(Q_{SOM}+Q_{RI})}\left(\frac{1}{2} - \frac{1}{1+r_{FM}e^{\frac{L}{\lambda_{Cu}}}}\right)\right]^{-1}, \quad (S23)$$

and interfacial contribution interface conductance ($G_I \neq 0$) and no bulk spin absorption:

$$\eta_{interfacial} = \frac{\Delta R_{NL}^{abs}}{\Delta R_{NL}^{ref}}\bigg|_{\substack{G_s^{SOM}=0 \\ Q_{SOM}=\infty}} = \left[1 + \frac{1}{(Q_\parallel+Q_{RI}/4)}\left(\frac{1}{2} - \frac{1}{1+r_{FM}e^{\frac{L}{\lambda_{Cu}}}}\right)\right]^{-1}. \quad (S24)$$

Finally, we assume the interfaces to be transparent as explained in the main text meaning that the interface conductance is $G_I = \infty$ and $Q_{RI} = 0$ leading to

$$\eta_{\text{bulk}} = \left[1 + \frac{1}{Q_{\text{SOM}}}\left(\frac{1}{2} - \frac{1}{1 + r_{\text{FM}}e^{\frac{L}{\lambda_{\text{Cu}}}}}\right)\right]^{-1}, \tag{S25}$$

$$\eta_{\text{interfacial}} = \left[1 + \frac{1}{Q_{\parallel}}\left(\frac{1}{2} - \frac{1}{1 + r_{\text{FM}}e^{\frac{L}{\lambda_{\text{Cu}}}}}\right)\right]^{-1}. \tag{S26}$$

Note that equation (S25) and (S26) are equivalent to equations (1) and (2) in the main text, respectively.

## S5. Spin properties of Py and Cu

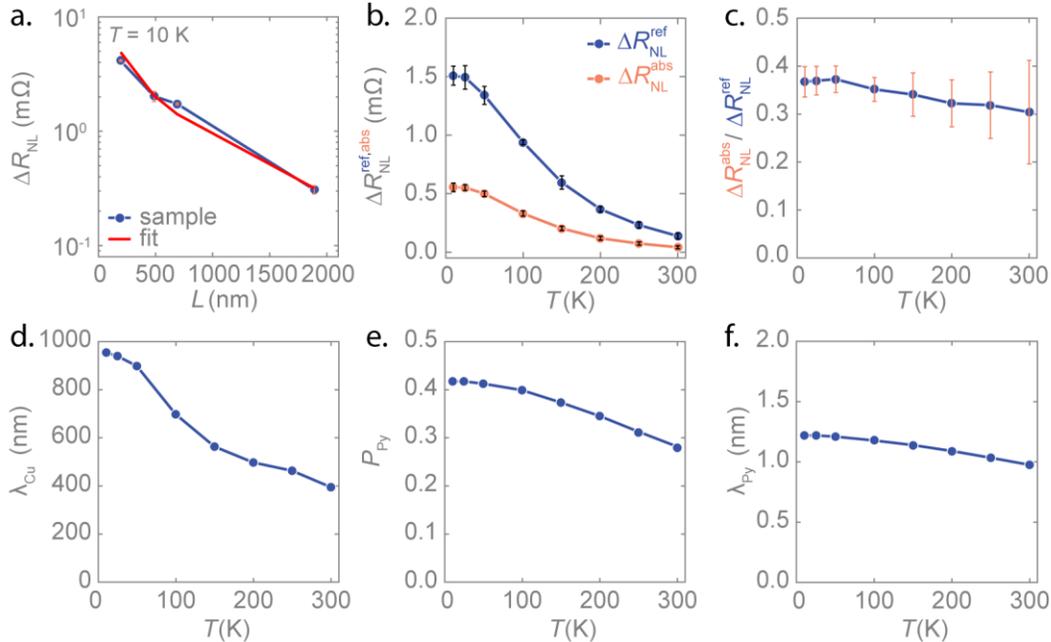

FIG. S5: *The spin properties of the Cu and Py electrodes. (a) The non-local spin signal in a Py/Cu LSV for various lengths of the Cu spin transport channel (i.e., the distance between the two Py electrodes). The red line is a fit with equation (S21). (b) The temperature dependence of the spin signal of a reference LSV ($\Delta R_{NL}^{ref}$) and the spin signal of a LSV with a middle Py electrode ($\Delta R_{NL}^{abs}$). (c) The ratio $\Delta R_{NL}^{abs}/\Delta R_{NL}^{ref}$ which can be fitted to equation (S22). Iterations of the fit with equation (S21) and equation (S22) result in the following spin properties: (d) the spin diffusion length of the Cu, (e) the spin polarization of the Py, and (f) the spin diffusion length of the Py.*

The spin properties of the Py and Cu channels need to be known to establish the spin properties of a W electrode or, in our case, an oxidized interface layer in a Py/Cu/W LSV. The protocol in Ref. [36] has been followed to acquire the spin properties of the Py and Cu channels. The analysis involves two types of LSVs. The first ones are several conventional Py/Cu LSV with different Py interelectrode distances $L$. Figure S5(a) plots the resulting spin signal as a function of $L$. A fit to equation (D21) can provide $\lambda_{\text{Cu}}$ and $P_{\text{Py}}$ by assuming a reasonable value for $\lambda_{\text{Py}}$. Secondly, the spin signal from a LSV with a middle electrode of Py is measured as shown in Fig. S5(b). The ratio of the two spin signals is presented in Fig. S5FIG.(c), which is used via equation (S22) to estimate $\lambda_{\text{Py}}$ for the $\lambda_{\text{Cu}}$ and $P_{\text{Py}}$ values extracted from the first experiment.

The final parameters, presented in Figs. S5(d), S5(e) and S5(f), are obtained by iterating this process several times, until a convergence of $\lambda_{Py}$, $\lambda_{Cu}$ and $P_{Py}$ is achieved. A Py/Cu interface resistance was measured but the spin resistance of the Py/Cu interface was not dominating, therefore we cannot separate the spin absorption of the bulk and the interface. $P_{Py}$ and $\lambda_{Py}$ are therefore effective values of the system and not necessarily representative of the Py electrode on its own. However, the spin parameters in Fig. S5 suffice for the purpose it has been used in this work, that is, extracting the spin properties of a middle SOM electrode.

## S6. Spin diffusion length for bulk spin absorption

An analysis on the spin diffusion length considering different spin absorption scenarios has been performed to identify where the spins are absorbed.

Scenario 1: Bulk spin absorption by the W electrode with a transparent Cu/W interface. Figure S6(a) displays $\lambda_W$ obtained by fitting equation (1) to $\Delta R_{NL}^{abs}/\Delta R_{NL}^{ref}$ and assuming $R_I = 0$. Once we know $R_I \neq 0$, this scenario cannot be applied. Note that the obtained spin diffusion length is comparable to reported values for Pt and Ta [16,27] and, hence, one should be careful to properly consider the interface.

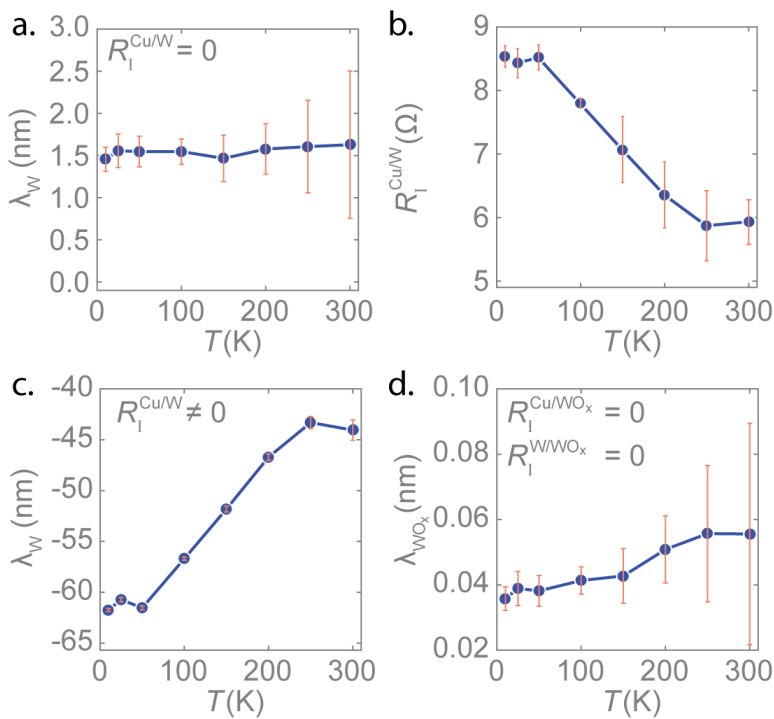

*FIG. S6: The spin diffusion length for three scenarios of bulk spin absorption in the Py/Cu/W LSV. The spin diffusion lengths are obtained from fitting different equations to the spin absorption data $\Delta R_{NL}^{abs}/\Delta R_{NL}^{ref}$ [see the inset of Fig. 2(d)]. The spin diffusion length is determined for (a) Spin absorption by the W electrode with a transparent Cu/W interface [$R_I^{Cu/W} = 0$, equation (1)]. (b) The interface resistance obtained with 3D FEM simulation considering a resistive interface with no thickness. (c) Spin absorption by the W electrode with a resistive Cu/W interface and no spin-loss (equation S23) where $R_I^{Cu/W}$ is presented in panel (b). (d) Spin absorption by the oxide layer with a transparent Cu/WO$_x$ and W/WO$_x$ interface ($R_I^{Cu/WO_x} = R_I^{W/WO_x} = 0$) and a resistivity $\rho_{WO_x}$ given in Fig. 1(d) of the main text [equation(1)].*

Scenario 2: Bulk spin absorption by the W electrode with an interface resistance at the Cu/W interface. If we did not uncover the interfacial oxide layer, the interface resistance measurements and 3D FEM simulations can be combined to extract an interface resistance. Figure S6(b) shows that these simulations give that this interface resistance in this case would be several Ω. In this scenario, if we assume no spin-loss at this interface, we can use equation (S23) which gives a negative $\lambda_W$ as presented in Fig. S6(c). A negative $\lambda_W$ is not a physical quantity for a spin diffusion length.

Scenario 3: Bulk spin absorption by the $WO_x$ layer with transparent $Cu/WO_x$ and $WO_x/W$ interfaces. The STEM images of the interface (Fig. 5) show an oxide layer between the Cu and W electrode. This interface layer is incorporated to the 3D FEM simulations resulting in a resistivity [Fig. 1(d)] that can be inserted in equation (1) to investigate spin absorption by the "bulk" oxide layer. The resulting $\lambda_{WO_x}$ is ~ 0.04 nm, which is shorter than the interatomic distance in transition metal oxides.

Therefore, we conclude that the spin absorption actually occurs at the $Cu/WO_x$ interface as presented in the main text.

## S7. Derivation of the spin-charge signal in a lateral spin valve

Non-local spin-charge technique is a frequently used tool to acquire the SCI efficiency of bulk SOM. This technique uses the same LSV structure as described in the spin absorption technique but with a different measurement configuration. Figure S7(a) displays the spin-to-charge measurement configuration. The injection of the spin current is identical to the spin injection in the LSV's explained in Supplemental material S4 but to measure the spin-to-charge conversion, the detector is the SOM middle electrode.

In this case, the spins are aligned by an external magnetic field ($H_x$) along the hard axis of the FM1 electrode ($x$-direction), a requirement for observing the SCI with this technique. The pure spin current is injected from FM1 electrode into the NM channel by $I_{bias}$. Subsequently, the spin current decays towards the SOM electrode, where it is partially absorbed. The spin current $I_{s,z}$ originating from this spin absorption is converted into a charge current $I_{SC}$ via spin-to-charge conversion. $I_{SC}$ is measured by detecting the transverse voltage $V_{SC}$ along the SOM electrode. The produced $I_{SC}$, and therefore $V_{SC}$, will revert when changing the FM1 magnetization with the external magnetic field. The difference in the spin-to-charge resistances ($R_{SC} = V_{SC}/I_{bias}$) for the two saturated magnetizations is twice the spin-to-charge signal $2\Delta R_{SC}$ as shown in Fig. S7(c).

The charge-to-spin conversion is discerned by swapping the injector and detector with respect to the spin-to-charge measurement, as presented in Figs. S7(b) and 3(a). A charge current is applied to SOM, in which the charge-to-spin conversion takes place, and a spin current $I_{s,-z}$ is created. This spin current will diffuse into the NM channel and travel towards the FM, that acts as the detector. The spin accumulation is probed as an open circuit voltage $V_{CS}$ across the FM/NM interface. The charge-to-spin resistance is defined as $R_{CS} = V_{CS}/I_{bias}$. Figure S7(c) shows that $R_{CS}$ as a function of $H_x$ changes sign compared to $R_{SC}(H_x)$ due to the swapping of the injector and the detector. Onsager reciprocity [37] dictates that $|\Delta R_{CS}| = |\Delta R_{SC}|$ as illustrated in Fig. S7(c).

In this analysis, bulk and interfacial SCI will be considered for the same reason as explained in Supplemental material S4. This means that the measured SCI signal $\Delta R_{SC}$ in Fig. S7(c) will contain contributions of both the bulk and interfacial SCI. Now, we derive the expression for $R_{SC}$. The charge current density induced by the inverse spin Hall effect, hence SCI in the bulk SOM, is given by

$$j_{c,y} = -\sigma_{\text{SOM}} \frac{\partial \mu_c^{\text{SOM}}}{\partial y} - \sigma_{\text{SH}} \frac{\partial \mu_s^{\text{SOM}}}{\partial z}, \tag{S27}$$

and the charge current density produced at the NM/SOM interface through interfacial SCI is

$$j_{c,\text{int}} = \sigma_{\text{SC}} \frac{\mu_c^{\text{SOM}} + \mu_c^{\text{NM}}}{2} \delta(z). \tag{S28}$$

$\sigma_{\text{SH}}$ is the spin Hall conductivity of the SOM and $\sigma_{\text{SC}}$ is the interfacial spin-charge conductivity (the equivalent of $\sigma_{\text{SH}}$ for the interface).

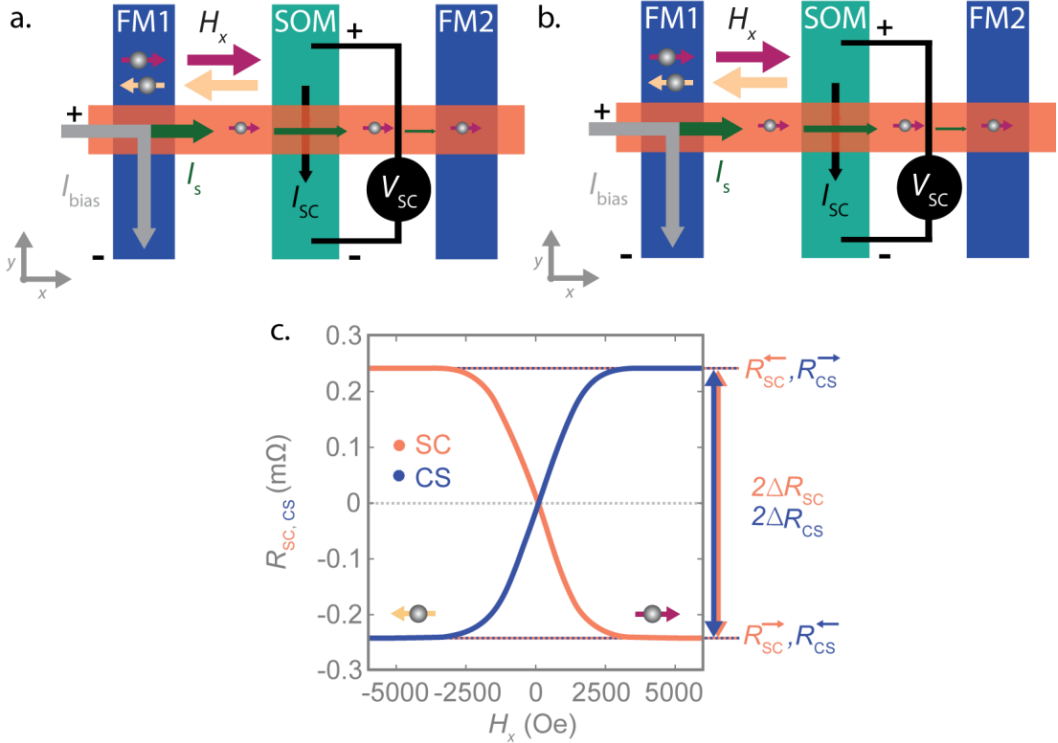

FIG. S7: *SCI measurements using the LSV. (a) The spin-to-charge (SC) measurement in the SOM electrode. A pure spin current is injected from the FM into the NM, that decays towards SOM, where it is partially absorbed. The spin current $I_{s,z}$ along $z$ with a polarization along $x$ is converted to a charge current $I_{SC}$ along $y$ and detected via a potential difference $V_{SC}$. An external magnetic field can reverse the magnetization of FM, consequently inverting the polarization of the spin current and reversing the direction of $I_{SC}$. (b) The charge-to-spin (CS) measurement has the injector and detector swapped in comparison to the SC configuration. By applying a charge current $I_{bias}$ through SOM, CS conversion in the SOM creates a spin accumulation at the NM/SOM interface. This spin accumulation will diffuse into the NM as a spin current $I_{s,x}$. The spin current is probed by the FM electrode giving $V_{CS}$. A reversal of the magnetization of FM reverses $\pm V_{CS}$. (c) The SC resistance ($R_{SC}$) and CS resistance ($R_{CS}$) as a function of the external magnetic field ($H_x$). The $R_{SC}$ transitions smoothly from a saturated negative $H_x$, to at saturated positive $H_x$. The gradual change is coming from the FM magnetization which undergoes a coherent rotation. $R_{CS}$ shows the reciprocal behavior due to the swapping of the current and voltage probes. The difference in $R_{SC}$ (or $R_{CS}$) between the saturated values is the spin-charge signal $2\Delta R_{SC} = 2\Delta R_{CS}$.*

The average current flowing in the SOM electrode is obtained by averaging over the cross section $A_{SOM} = t_{SOM} w_{SOM}$. This current should vanish because the experiment probes a voltage in open-circuit conditions and, therefore,

$$0 = -\sigma_{SOM} \frac{\partial \mu_c^{SOM}}{\partial y} + \sigma_{SH} \frac{1}{t_{SOM}} \mu_s^{SOM}(0) + \sigma_{SC} \frac{1}{t_{SOM}} \frac{\mu_s^{SOM} + \mu_s^{NM}}{2}\bigg|_{z=0}, \quad (S29)$$

and integration over $y$ gives

$$eV_{SC} = -\mu_c^{SOM}\bigg|_{-\frac{L_y}{2}}^{+\frac{L_y}{2}} 0 = -\frac{\sigma_{SH}}{\sigma_{SOM}} \frac{w_{NM}}{t_{SOM}} \mu_s^{SOM}(0) - \frac{\sigma_{SC}}{\sigma_{SOM}} \frac{w_{NM}}{t_{SOM}} \frac{\mu_s^{SOM} + \mu_s^{NM}}{2}\bigg|_{\substack{z=0 \\ x=0}}. \quad (S30)$$

The spin-charge resistance is defined as

$$R_{SC} = \frac{V_{SC}}{J_{bias} A_{int}^{FM}}, \quad (S31)$$

such that, combining equation (S30) and (S31) results in

$$R_{SC} = \frac{w_{NM} R_s^{FM}}{t_{SOM} \sigma_{SOM}} \frac{4P e^{\frac{L}{2\lambda_{NM}}}[2Q_\parallel \sigma_{SC}(Q_{RI} + 2Q_{SOM}) + Q_{SOM} \sigma_{SH}(4Q_\parallel - Q_{RI})]}{r_{FM} e^{\frac{L}{\lambda_{NM}}}[4Q_\parallel r + 2(Q_{RI} + 2)Q_{SOM} + Q_{RI}] + 4Q_\parallel(r-2) - 2(2-2Q_{RI})Q_{SOM} + Q_{RI}}. \quad (S32)$$

Note that equation (S32) is derived considering the presence of the second FM electrode, as presented in the spin absorption technique [Figs. S7(a) and S7(b)].

The change of polarization of the spin current, when reversing the magnetization of the FM injector, induces a detectable change in $R_{SC}$. The difference in $R_{SC}$ between the saturated positive and negative magnetization of the FM is twice the SCI signal, $2\Delta R_{SC} = R_{SC}^{\rightarrow} - R_{SC}^{\leftarrow}$.

The 1D spin diffusion model gives equation (S32) that explains the theoretical value of the SCI signal in the LSV devices. However, the electrical shunting $x_{SOM}$ of the current produced in the SOM electrode by the NM channel is not included in the model. When this electrical shunting is not considered, the spin-charge conductivity will be underestimated. Therefore, $x_{SOM}$ is obtained by 3D finite element method (FEM) simulations (see Supplemental material S8) such that, finally,

$$\Delta R_{SC} = x_{SOM} R_{SC}. \quad (S33)$$

Similar to the spin absorption as explained in Supplemental material S4, the SCI also has the two limiting cases:

bulk SCI including an interface conductance ($G_I \neq 0$) but no interfacial spin-loss:

$$\Delta R_{SC}\big|_{\substack{G_\parallel = 0 \\ Q_\parallel = \infty}} = x_{SOM} \frac{w_{NM} R_s^{FM}}{t_{SOM} \sigma_{SOM}} \frac{4P Q_{SOM} \sigma_{SH} e^{\frac{L}{2\lambda_{NM}}}}{r_{FM}^2 r_{SOM} e^{\frac{L}{\lambda_{NM}}} + r_{SOM} - 2}. \quad (S34)$$

and interfacial SCI with interface conductance ($G_I \neq 0$) and no bulk SCI:

$$\Delta R_{SC}\big|_{\substack{G_s^{SOM}=0 \\ Q_{SOM}=\infty}} = x_{SOM} \frac{w_{NM} R_s^{FM}}{t_{SOM}\sigma_{SOM}} \frac{4PQ_\parallel \sigma_{SC} e^{\frac{L}{2\lambda_{NM}}}}{r_{SOM}^2 r_{RI} e^{\frac{L}{\lambda_{NM}}} + r_{RI} - 2},\tag{S35}$$

given that $r_{RI} = 1 + 2Q_\parallel + Q_{RI}/2$.

For a transparent interface, that is, $G_I = \infty$ and $Q_{RI} = 0$, we obtain:

Bulk SCI:

$$\Delta R_{SC}\big|_{\substack{G_\parallel=0 \\ Q_\parallel=\infty}} = x_{SOM} \frac{w_{NM} R_s^{FM}}{t_{SOM}\sigma_{SOM}} \frac{4PQ_{SOM} \sigma_{SH} e^{\frac{L}{2\lambda_{NM}}}}{r_{FM}^2 (1+2Q_{SOM}) e^{\frac{L}{\lambda_{NM}}} + 2Q_{SOM} - 1}.\tag{S36}$$

And interfacial SCI:

$$\Delta R_{SC}\big|_{\substack{G_s^{SOM}=0 \\ Q_{SOM}=\infty}} = x_{SOM} \frac{w_{NM} R_s^{FM}}{t_{SOM}\sigma_{SOM}} \frac{4PQ_\parallel \sigma_{SC} e^{\frac{L}{2\lambda_{NM}}}}{r_{FM}^2 (1+2Q_\parallel) e^{\frac{L}{\lambda_{NM}}} + 2Q_\parallel - 1}.\tag{S37}$$

Equation (S37) is equivalent to equation 3 in the main text.

## S8. Shunting factor in the Cu/WOₓ/W heterostructure using a 3D FEM simulation

Electrical shunting is essential for the evaluation of SCI properties of the SOM or spin-orbit interface in LSVs. Electrical shunting is the effect of electrical current flowing through the lowest-resistance path by passing around different components in the circuit. More specifically, in LSVs this means that the ISHE current ($I_{ISHE}$) generating the ISHE voltage ($V_{ISHE}$) is partially shunted by the Cu channel connected to the SOM electrode. Figure S8(a) shows an illustration of spin-to-charge conversion at the interface region in a LSV where the ISHE voltage is probed at the two ends of the SOM electrode in open-circuit conditions. In this condition the $I_{ISHE}$ flows back into the NM as $I_{NM}$ and the SOM as $I_{SOM}$ [38].

Shunting in a simple one-dimensional two-resistor model gives that $I_{SOM} = I_{ISHE}/x_{SOM}$ where $x_{SOM}$ is the shunting factor of SOM. Consequently, $V_{ISHE} = R_{SOM} I_{SOM} = R_{SOM} I_{ISHE}/x_{SOM}$. Even though the one-dimensional model with two electrodes as independent parallel resistors helps to understand the shunting factor, it fails to exactly reproduce the electrical shunting while in the real device the two electrodes are connected at every point of the interface. 3D FEM simulations can overcome this problem as an electrical connection at the overall interface of the SOM and NM electrodes can be included by setting proper boundary conditions. The 3D FEM simulation improves the estimation of $x_{SOM}$ of $I_{ISHE}$ in the LSV and, therefore, allows determining more precisely the SCI parameters.

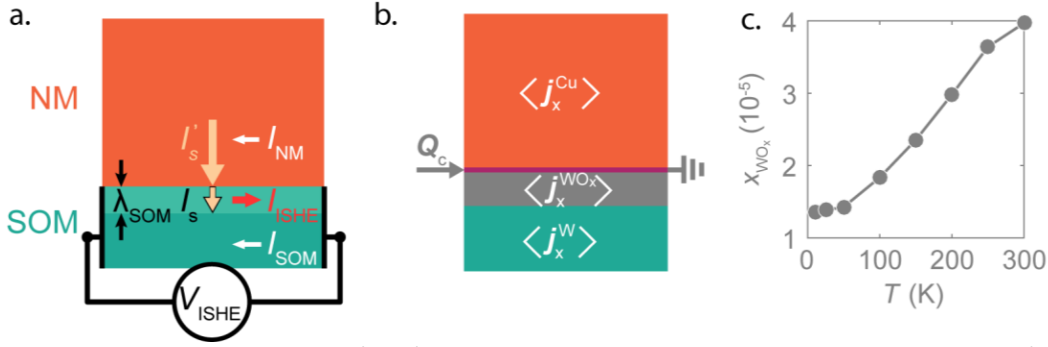

FIG. S8: Electrical shunting in the Cu/WOx/W heterostructure. (a) An illustration of the ISHE at the NM/SOM interface region of a LSV where the spin absorption and spin-to-charge current conversion takes place. The current $I_{ISHE}$ is generated in the SOM over a distance defined by the spin diffusion length $\lambda_{SOM}$. The $V_{ISHE}$ created by $I_{ISHE}$ is probed in open-circuit conditions resulting in the flow of $I_{NM}$ and $I_{SOM}$ in the opposite direction as $I_{ISHE}$. This figure is adapted from Ref. [38]. (b) In the 3D FEM simulation we apply a line current source $Q_C$ to the Cu/WOx interface and extract the average current density $\langle j_x^i \rangle$ in each component; (c) The electrical shunting factor $x_{WO_x}$ calculated with $\langle j_x^{WO_x} \rangle$ from the 3D FEM simulation and $x_{WO_x} = \langle I_{ISHE} \rangle / \langle I_{WO_x} \rangle$ for the interfacial oxide layer.

We used *Comsol Multiphysics* to perform the 3D FEM simulation and extract the electrical shunting, that is the fraction of the $I_{ISHE}$ flowing through each layer in the Cu/WOx/W structure. The additional oxide layer makes the geometry in the 3D FEM a tri-layer of Cu, WOx and W as shown in Fig. S8(b). The electrical shunting is attained by simulating the average current density $\langle j_x^i \rangle$ with $i$ being Cu, WOx and W from where the current flow $\langle I_i \rangle \, (= \langle j_x^i \rangle / A_i)$ in each component is calculated. The SCI occurs at the Cu/WOx interface, thus $I_{ISHE}$ is generated at this interface. The voltage drops over the three components (Cu, WOx and W) is the same according to Kirchhoff's law. However, equation (3) considers the current flow in the WOx, indicated by the appearance of $t_{WO_x}$ and $\sigma_{WO_x}$, such that, even though $V_{ISHE}$ will be probed at the W electrode, the electrical shunting factor that has to be considered is the one of the oxide, that is, $x_{WO_x}$.

The generation of $I_{ISHE}$ at the Cu/WOx interface brings about another discussion which is where to apply the charge current density in the 3D FEM simulation. Following the 1D model for electrical circuits, the charge current density $j_C$ [A/m$^2$] should be applied to the oxide layer. However, as we have found that the origin of $I_{ISHE}$ is at the Cu/WOx interface, we must adapt the model and apply $Q_C$ [A/m] at the interface itself. We obtain $x_{WO_x}$ for a temperature range of 10-300 K as displayed in Fig. S8(c). The change in $x_{WO_x}$ between applying $j_C$ (not shown) and $Q_C$ is roughly two orders of magnitude. Such a big change appears in the Cu/WOx/W structure because the high resistivity of the oxide layer plays a dominant role in the structure. Note that $I_{ISHE}$ in metallic NM/SOM structures is created in a region close to the interface with a thickness defined by $\lambda_{SOM}$, therefore, the $j_C$ imitating $I_{ISHE}$ should be applied to this region. However, generally, NM and SOM have a "low" resistivity and no significant change in $x$ for the two configurations will be observed. Finally, $x_{WO_x}$ is used to quantify the SCI in the Cu/WOx interface.